\documentclass[structabstract]{aa-package/aa}  
\usepackage{color}
\usepackage{graphicx}
\usepackage{aa-package/bibtex/natbib}
\usepackage[varg]{txfonts}
\usepackage{longtable,lscape}
\usepackage{url}
\usepackage{verbatim}
\usepackage{dcolumn}

\usepackage{txfonts}
\newcolumntype{d}[1]{D{.}{\cdot}{#1}}

\newcolumntype{.}{D{.}{.}{-1}}

\newcommand{\nsources}{6639} 
\newcommand{\nmsx}{2314}
\newcommand{\niras}{755} 
\newcommand{\nall}{645}
 
\newcommand{\nnothing}{4216}
\newcommand{\area}{153}
\newcommand{\sex}{\texttt{SExtractor}}
\newcommand{\rms}{r.m.s.}

\begin{document}
\bibliographystyle{aa-package/bibtex/aa}
\title{ATLASGAL --- Compact source catalogue: 330\degr\ $ <\ell <$ 21\degr \thanks{The full catalogue and the calibrated emission maps are only available in electronic form at the CDS via anonymous ftp to cdsarc.u-strasbg.fr (130.79.125.5) or via http://cdsweb.u-strasbg.fr/cgi-bin/qcat?J/A+A/.} }
\authorrunning{Contreras et al.}
\titlerunning{Compact source catalogue}

\author { Y.\,Contreras \inst{1,2}\thanks{Current address: CSIRO Astronomy and Space Science, P.O. Box 76, Epping NSW 1710, Australia. \email{yanett@das.uchile.cl}}, F.\,Schuller\inst{1,3},
  J.\,S.\,Urquhart\inst{1}, T.\,Csengeri\inst{1}, F.\,Wyrowski\inst{1}, H.\,Beuther\inst{4}, S.\,Bontemps\inst{5}, L.\,Bronfman\inst{2}, T.\,Henning\inst{4}, K.\,M.\,Menten\inst{1}, P.\,Schilke\inst{6}, C.\,M.\,Walmsley\inst{7,8}, M.\,Wienen\inst{1}, J.\,Tackenberg\inst{4}, H.\,Linz\inst{4} 
  }

  \institute { Max-Planck-Institut f\"ur Radioastronomie, Auf dem H\"ugel 69,
    Bonn, Germany
    \and Departamento de Astronom\'{i}a, Universidad de Chile, Casilla 36-D, Santiago, Chile
    \and European Southern Observatory, Alonso de Cordova 3107, Vitacura, Santiago, Chile
    \and Max-Planck-Institut f\"ur Astronomie, K\"onigstuhl 17, 69117 Heidelberg, Germany 
    \and Laboratoire d'Astrophysique de Bordeaux –- UMR 5804, CNRS – Universit\'e Bordeaux 1,
    BP 89, 33270 Floirac, France 
    \and I. Physikalisches Institut, Universit\"at zu K\"oln, Z\"ulpicher Str. 77, 50937 K\"oln, Germany 
    \and Osservatorio Astrofisico di Arcetri, Largo E. Fermi, 5, 50125 Firenze, Italy
    \and Dublin Institute for Advanced Studies, Burlington Road 10, Dublin 4, Ireland
}

\date{Received xxx; accepted xxx}

% \abstract{}{}{}{}{} 
% 5 {} token are mandatory

\abstract
{The APEX Telescope Large Area Survey of the Galaxy (ATLASGAL) is the
  first systematic survey of the inner Galactic plane in the
  sub-millimetre. The observations were carried out with the Large APEX
  Bolometer Camera (LABOCA), an array of 295 bolometers
  observing at 870~$\mu$m (345 GHz).  }
{ Here we present a first version of the compact source catalogue extracted from
  this survey. This catalogue provides an unbiased database of dusty clumps in
  the inner Galaxy.  }
{ The construction of this catalogue was made using the source
 extraction routine \sex. We have cross-associated the obtained
  sources with the IRAS and MSX catalogues, in order to constrain their nature.}
{ We have detected \nsources\ compact sources in the range from $330
  \leq \ell \leq 21$ degrees and $|b| \leq 1.5$ degrees. The
    catalogue has a 99\% completeness for sources with a peak flux above
    6$\sigma$, which corresponds to a flux density of
    $\sim$0.4\,Jy\,beam$^{-1}$. The parameters extracted for sources
    with peak fluxes below the 6$\sigma$ completeness threshold should
    be used with caution. Tests on simulated data find the uncertainty
    in the flux measurement to be $\sim$12\%, however, in more complex
    regions the flux values can be overestimated by a factor of 2 due
    to the additional background emission. Using a search radius of
    30\arcsec\ we found that 40\% of ATLASGAL compact sources are
    associated with an IRAS or MSX point source, but, $\sim$50\% are
    found to be associated with MSX 21 $\mu$m fluxes above the local
    background level, which is probably a lower limit to the actual number of
    sources associated with star formation.} 
{  Although infrared emission is found towards the majority of the
  clumps detected, this catalogue is still likely to include a
  significant number of clumps that are devoid of star formation
  activity and therefore excellent candidates for objects in the
  coldest, earliest stages of (high-mass) star formation.}  
\maketitle
% 
% ________________________________________________________________

\section{Introduction}

The processes and time scales involved in the formation of high-mass stars are
still poorly understood, although such stars play an important role in the
evolution of their parental clouds and the subsequent star formation inside
them \citep{krumholz-2007}. Moreover, they interact with their environment
injecting an enormous amount of energy both radiative and mechanical
\citep{hoare-2007}.

Great advances have been made in understanding how low-mass stars form
\citep{mckee-ostriker-2007}, but the situation for high-mass stars is still
very unclear. Different scenarios have been proposed that provide a
theoretical framework for the formation of high-mass stars such as competitive
accretion \citep{bonnell-2004} and core accretion \citep{mckee-tan-2003}. On
the observational side, the short time scales \citep{mckee-tan-2002} involved
in the different stages of the formation process make it difficult to build
large representative samples for the statistical studies necessary to robustly
test these theories.
 
Several surveys have been carried out to study the environments of high-mass
star-forming regions. The drawback of many of these surveys is that they
target samples that fulfil certain selection criteria (e.g. maser emission,
or peculiar far-infrared colours), and, as such, they are biased towards given
evolutionary stages (see e.g. \citealp{wyrowski-2007}). For example, methanol
masers have been found to be exclusively associated with the very early stages
of high-mass star formation up to the more evolved ultracompact HII region
phase and were the subject of the recent methanol multibeam (MMB;
\citealt{green+2009}) survey. The Red MSX Source survey (RMS;
\citealt{urquhart2007c}) targets high-mass young stellar objects (MYSOs) that
are known to be mid-infrared bright, while the {\bf{C}}o-{\bf{O}}rdinated
{\bf{R}}adio a{\bf{n}}d {\bf{I}}nfrared {\bf{S}}ource for {\bf{H}}igh-Mass
Star Formation (CORNISH; \citealt{hoare2012}) targets the
ultra-compact (UC) {\sc Hii} region stage at which point the young embedded star begins
to ionize its surrounding envelope and hence represents a later evolutionary
stage.

All of these surveys provide useful statistics on the particular stage
targeted (an evolutionary snapshot) but do not provide a complete
overview of high-mass star formation. What is needed is a Galaxy wide
survey using an unbiased tracer that links all of these complementary
surveys together. The ideal tracer to choose is the submillimetre
thermal emission from dust as it is optically thin, and so traces the
whole column of material across the Galaxy, and is associated with all
stages of high-mass star formation. However, until recently,
limitations in mapping speeds have resulted in limited coverage of the
Galactic plane. For example, SCUBAs\footnote{Submillimeter Common User
  Bolometer Array, \citep{Holland}} 9-year lifetime, a total area of
only 29.3 deg$^2$ has been surveyed at 850 $\mu$m
\citep{di_francesco2008}.

\begin{figure*}
\begin{center}
\includegraphics[width=0.49\textwidth, trim= 0 0 0 0]{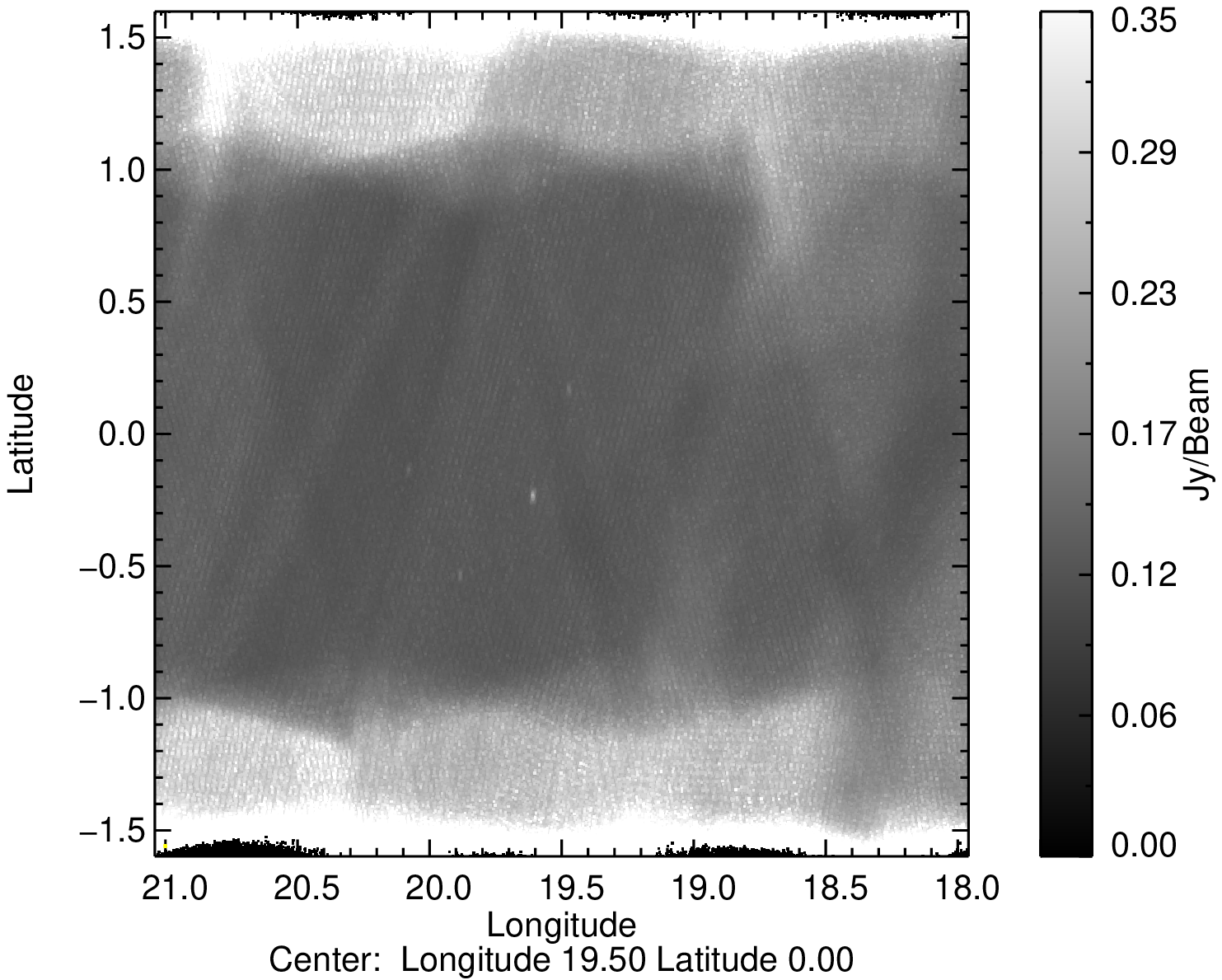}
\includegraphics[width=0.47\textwidth, trim= 0 0 0 0]{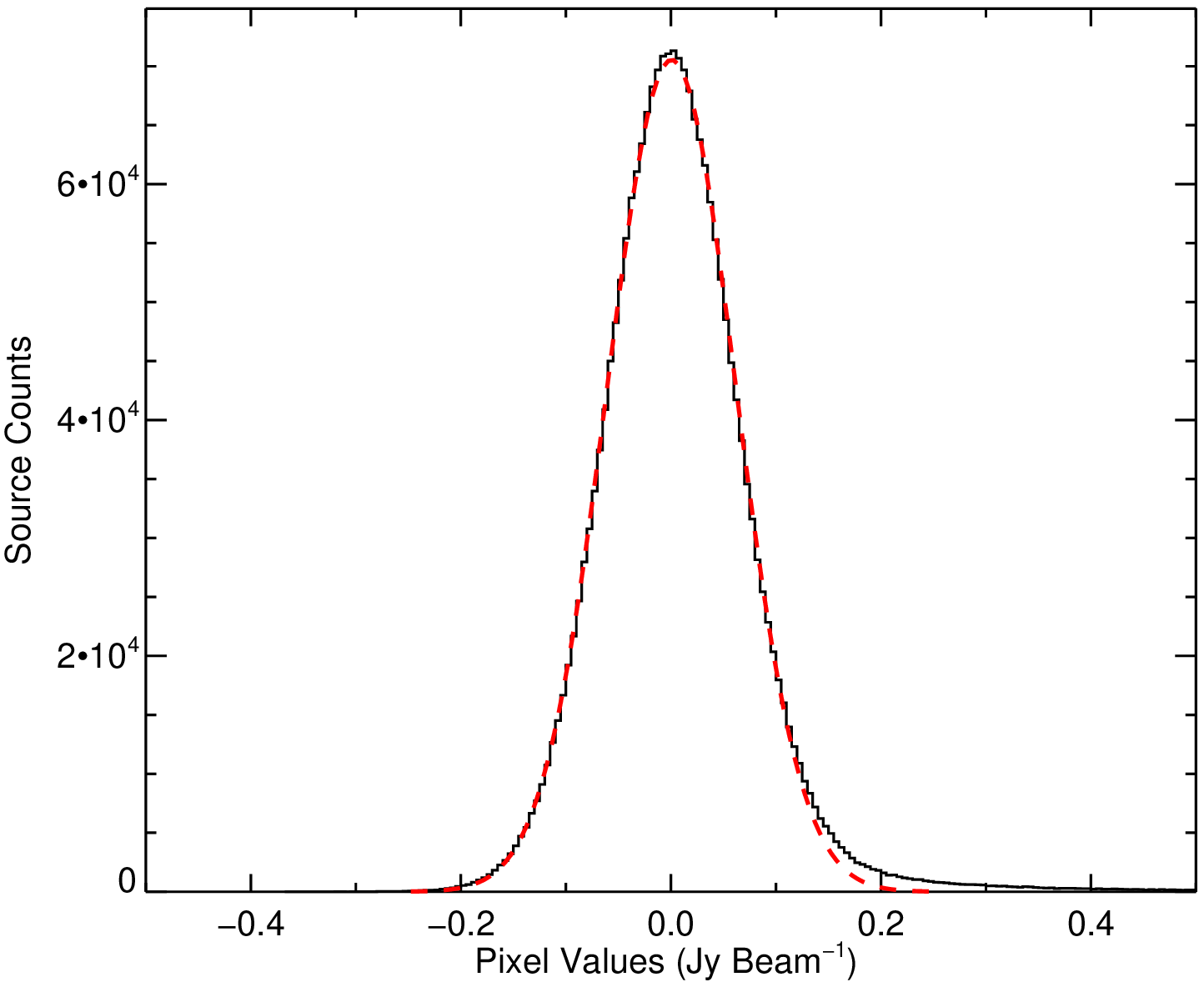}

\caption{\label{fig:noise_map} In the left panel we present an \rms\ noise map
  of a 3\degr$\times$3\degr\ field centred on $\ell=19.5\degr$ and $b=0$
  showing how the noise varies as a function of Galactic longitude and
  latitude. In the right panel we present a histogram of the pixel values
  found within $|b| < 1\degr$ where the noise is approximately constant. The
  red line show the result of a Gaussian fit to the noise, from which we
  estimate the average noise for this field
  ($\sigma_{\rm{r.m.s.}}\simeq60$\,mJy). }

\end{center}
\end{figure*}

The need for unbiased surveys to get a complete census of the regions where
high-mass stars are being formed has led to several projects to map the
Galactic plane at millimetre and submillimetre wavelengths. The APEX\footnote{This publication is based on data acquired with the Atacama
  Pathfinder Experiment (APEX) is a collaboration between the
  Max-Planck-Institut f\"ur Radioastronomie, the European Southern
  Observatory, and the Onsala Space Observatory.} Telescope Large Area Survey
of the Galaxy \citep[ATLASGAL;][]{Schuller} is the first complete survey of the inner Galactic plane providing an unbiased view and complete census of high-mass star-forming cores and clumps. It is well complemented by the Herschel Infrared Galactic plane \citep[Hi-GAL;][]{higal} survey in the far-infrared range at a similar high spatial resolution of $\sim$20\arcsec\ (e.g., 0.5\,pc at 5\,kpc), and at slightly longer wavelengths with the Bolocam Galactic Plane Survey \citep[BGPS;][]{aguirre2011}. Sub-mm surveys, as opposed to other surveys at
shorter wavelengths such as the Galactic Legacy Infrared Mid-Plane Survey
Extraordinaire \citep[GLIMPSE][]{benjamin-2003} or the MIPS Inner Galactic
Plane Survey \citep[MIPSGAL;][]{carey-2009}, are sensitive to dust in a broad
range of temperatures, including the coldest dust.  Therefore, such surveys
have the potential to trace the earliest stages of star formation.

In this paper, we present a compact source catalogue extracted from a
limited range of the ATLASGAL survey.  ATLASGAL is an unbiased survey
of the Galactic plane at 870\,$\mu$m with LABOCA \citep[Large APEX Bolometer
Camera;][]{siringo-2009}, and is the largest and most sensitive ground-based sub-mm
survey to date. The catalogue presented here is extracted from a
subset of the data collected from 2007 to 2010, which covered \area\
deg$^2$, about one third of the total survey area (420
deg$^2$). A catalogue will be compiled from the remaining fields and be published in a companion paper when the data reduction and analysis of these fields has been completed.

The aim of this survey is to obtain a complete census of cold dusty clumps in
the Galaxy, and to study their distribution across the Galaxy. This survey
reveals a complete sample of different stages of high-mass star formation. It
will be complementary to the other Galactic plane surveys such as UKIDSS
(\citealt{ukidss2007}) and {\bf{V}}ISTA {\bf{V}}ariables in the {\bf{V}}ia
Lactea (VVV; \citealt{vvv2012}) at the near-infrared; GLIMPSE and MIPSGAL at
the mid- to far-infrared (\citealt{benjamin-2003} and \citealt{carey-2009}
respectively); Hi-GAL which spans the far-infrared and sub-millimetre
wavelengths (70-500\,$\mu$m; \citealt{higal}) between MIPSGAL and ATLASGAL;
and finally the 5\,GHz CORNISH Survey. These surveys combined together will
provide a complete wavelength coverage of the Galactic plane from the
near-infrared to the radio, and will also form a unique database for follow-up
observations.

To demonstrate the potential impact of this unique database, the
ATLASGAL data has been used in numerous studies; \citet{wienen2012} determined the temperatures and kinematic
distances of $\sim$1000 clumps in various evolutionary stages using
ammonia follow-up observations; $\sim$1500 ATLASGAL sources are part
of the MALT90 project
\citep{Foster-2011}\footnote{http://malt90.bu.edu/}, which
simultaneously maps these sources in 16 spectral lines in the 3-mm
wavelength range with the Mopra radio telescope;
\citet{deharveng+2010} studied the association of infrared bubbles
with ATLASGAL dust clumps; and based on ATLASGAL data
\citet{Tackenberg-2012} identified a number of starless cores, using
indicators of star formation in GLIMPSE and MIPSGAL images.

In Sect.\,\ref{obs}, we describe the observations, and in Sect.\,\ref{sex} we
detail how the sources were extracted. Tests on the reliability of the
extraction algorithm are described in Sect.\,\ref{art}. In Sect.\,\ref{cat} we
present the compact source catalogue and discuss the distributions of the
various source parameters, while in Sect.\,\ref{cross} we investigate the
correlation of ATLASGAL sources with the IRAS and MSX source catalogues and
with other types of objects. Finally, we summarise our results in
Sect.\,\ref{sum}.

% _________________________________________________________________

\section{Observations}
\label{obs}

The observations were carried out with LABOCA \citep{siringo-2009} at the APEX
(Atacama Pathfinder EXperiment) 12\,m submillimetre antenna \citep{guesten},
located in Llano de Chajnantor, Chile.  The typical pointing \rms\ error was
measured to be $\sim$\,4$''$.  The bandpass of LABOCA is centred at 345\,GHz,
with a bandwidth of 60\,GHz. At this frequency, the APEX beam FWHM is 
$19{\rlap.{}''}2$.

In the present paper, we will focus on the region $330\degr \leq \ell \leq
21\degr$, $-1.5\degr \leq b \leq 1.5\degr$ (\area\ deg$^2$). Future versions
of the catalogue will cover the full 420 deg$^2$ observed as part of
ATLASGAL. 

The observations were made using on-the-fly maps with a scanning
velocity of 3$'/s$.  They produced slices of 1$\degr$ width in
Galactic longitude times 2$\degr$ length in Galactic latitude for the
observations carried out during 2007 and slices of 1$\degr$ in
Galactic longitude times 2 to 3$\degr$ in Galactic latitude during
2008 to 2010. To avoid artefacts inherent to the observing technique,
each position in the sky is covered at least twice, in two or more
maps observed with different scanning angles. The flux calibration was
made using planets as primary calibrators (Mars, Neptune and Uranus),
and secondary calibrators (bright Galactic sources with known fluxes
from the commissioning of LABOCA). The errors in peak flux are
estimated to be lower than 15\% \citep{Schuller}.

The data reduction was made using the BOlometer array data Analysis package
\citep[BoA;][]{Schuller2012}. The steps involved in the data reduction are
explained in detail in \citet{Schuller}, and can be summarised as: flagging
bad bolometers, removing the correlated noise, flagging noisy bolometers,
despiking, low frequency filtering, and subtracting a first order
baseline. The maps are then built by co-adding the signals from all
bolometers, using a weighted mean with natural weighting. A weight map is also
calculated from the sum of all the weights that contribute to each pixel in the final map. An \rms\ map can then be computed as 1/sqrt(weight).

A problem inherent to the reduction of ground-based bo\-lo\-meter data is the
loss of the extended emission, which is filtered out when the correlated noise
is subtracted. As a result, uniform emission at angular scales larger than
$2{\rlap.{}'}5$ is removed. This reduction scheme was optimised to recover
compact sources.  The final maps are built with a pixel size of 6\arcsec,
corresponding to a $\sim$1/3 of the telescope beam. The size of each map is of
$3\times3$ degree, having an overlap between two consecutive maps of $\sim$4.5
arc minutes.

\begin{figure}
\begin{center}

\includegraphics[width=0.49\textwidth, trim= 0 0 0 0]{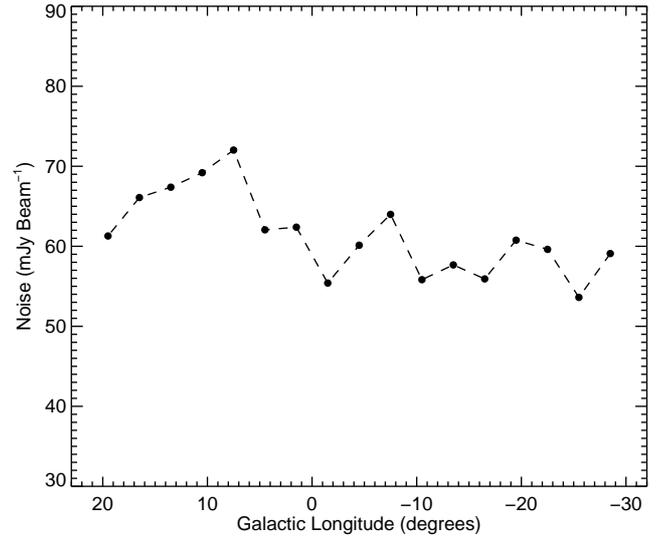}

\caption{\label{fig:field_noise_map} Plot showing how the noise varies as a function of Galactic longitude. The points show the standard deviation of the pixel to pixel variation averaged for each 3\degr$\times$3\degr\ field, however, we have only included pixels with $|b| < 1\degr$ where the noise is relatively uniform and where the vast majority of the 870\,$\mu$m emission is concentrated. A dashed line connects the point to emphasis the field-by-field variations in the noise.} 

\end{center}
\end{figure}

In the left panel of Fig.\,\ref{fig:noise_map} we present an example
of a final noise map, i. e. the \rms\ noise level distribution over a
reduced map. As can be seen from this map the noise is reasonably
uniform within $|b| < 1\degr$, however, due to poorer coverage at
higher latitude ($|b| > 1\degr$) the pixel to pixel \rms\ is
significantly larger, typically 150-250\,mJy\,beam$^{-1}$. The vast
majority of the emission seen in the ATLASGAL maps is concentrated
close to the Galactic mid-plane and so the sensitivity of the survey
can be considered fairly uniform. In the right panel of
Fig.\,\ref{fig:noise_map} we present a histogram of the flux values of
pixels with $|b| < 1\degr$. The \rms\ noise is well modelled by a
Gaussian fit (shown as a dashed red line), from which we estimate the
\rms\ noise to be $\sim$\,60\,mJy\,beam$^{-1}$. We estimate the \rms\
noise in each field in the same way and present a plot showing how the
noise varies as a function of Galactic longitude in
Fig.\,\ref{fig:field_noise_map}. This plot reveals that the noise is
fairly uniform for the regions presented here
($\sim$\,50-70\,mJy\,beam$^{-1}$), with the observed variations
resulting from different weather conditions and elevation of the
various observations. This is particular noticeable for the maps
  within +7 and +17 degrees where the higher noise in these maps is
  believed to be due a combination of high airmass and poorer weather
  condition.

% __________________________________________________________________
\section{Source extraction}
\label{sex}
\begin{figure*}
\begin{center}
\includegraphics[width=0.49\textwidth, trim= 0 0 0 0]{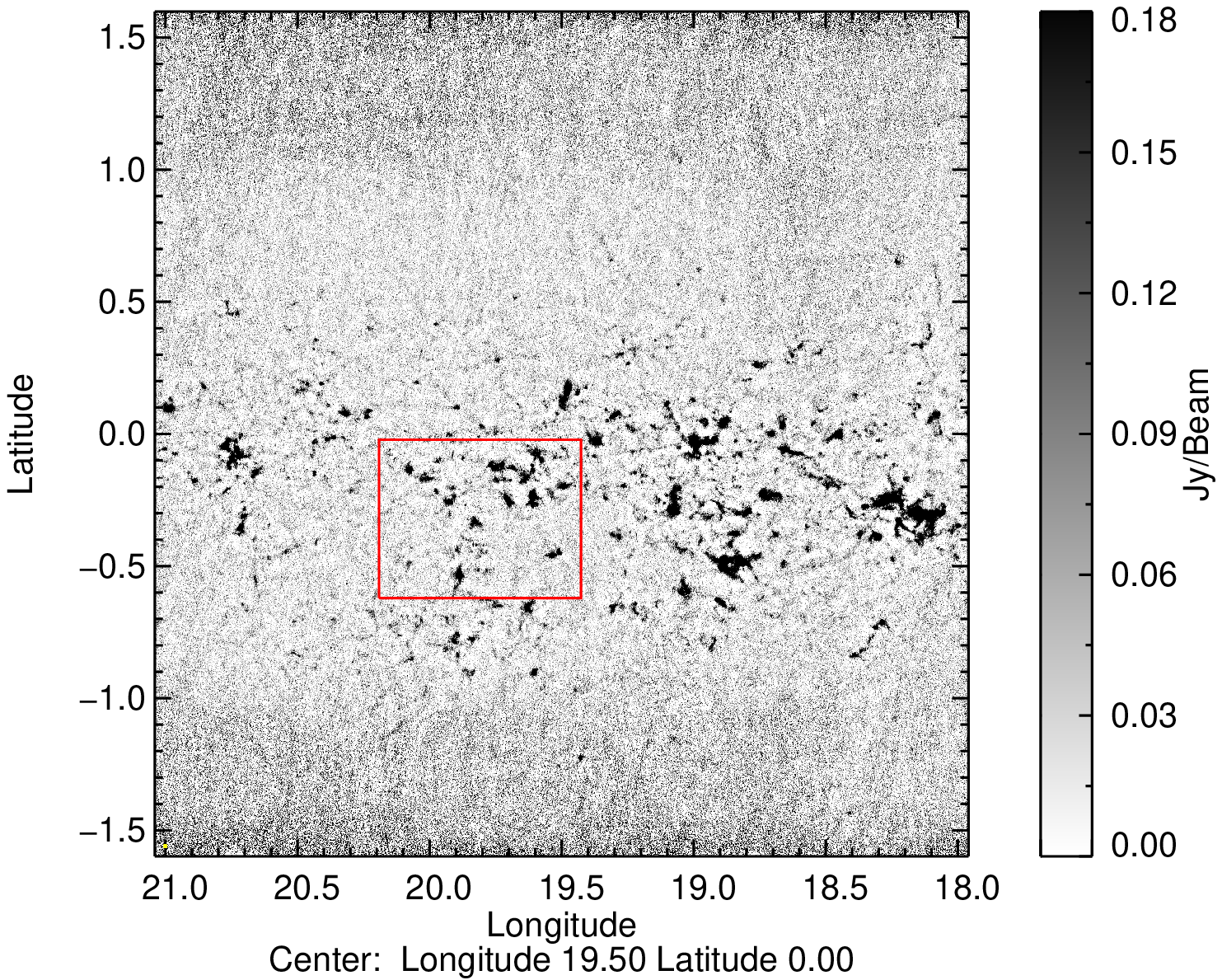}
\includegraphics[width=0.49\textwidth, trim= 0 0 0 0]{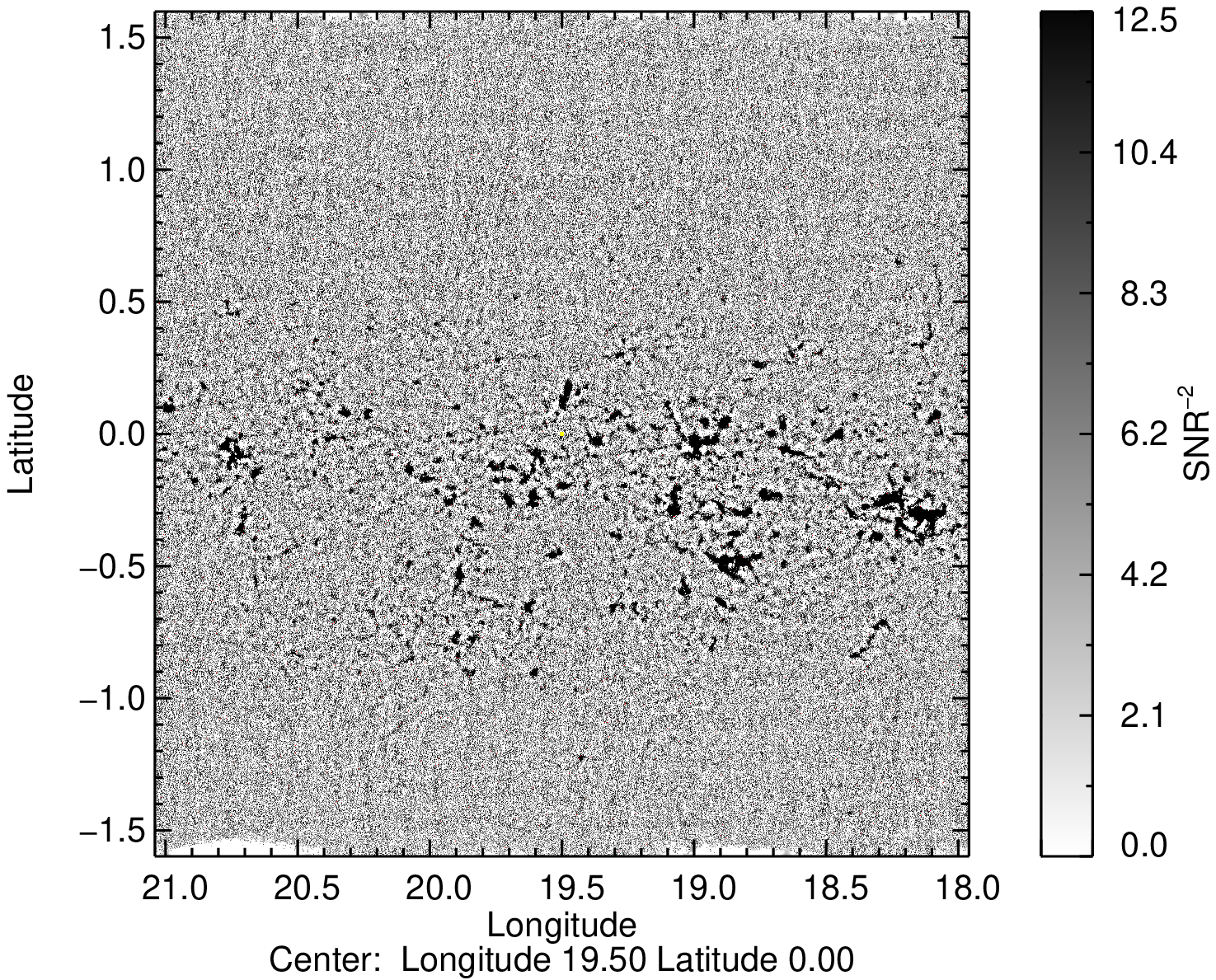}

\caption{\label{fig:emission_snr_maps} Example of an ATLASGAL emission map (left panel) and the corresponding filtered signal-to-noise ratio map (see text for details). The region outlined in red indicates the field presented in Fig.\,\ref{fig:example_detection}.} 

\end{center}
\end{figure*}

To create the compact source catalogue we used the source extraction algorithm
\sex\ \citep{bertin}. Although \sex\ was originally created as a method of
source extraction for visible and infrared images, it has been used with good
results for the extraction of sub-mm sources as well by \citet{scuba_thompson}
and more recently by the \emph{Planck} team to produce the higher frequency
components of their compact source catalogue (217-857\,GHz;
\citealt{planck2011}).

Most source extraction algorithms assume that the noise level is constant
across the emission map supplied. However, in most surveys this is unlikely to
be the case due to variations in coverage, weather, elevation and instrumental
effects, which make the noise level vary across the field, and this is indeed
the case for the ATLASGAL survey. In Fig.\,\ref{fig:noise_map} we present the
noise map for the $3\degr\times3\degr$ field centred on $\ell=19.5\degr$ and
$b=0$. The fact that most source extraction algorithms assume a constant noise
can lead to real sources lying in areas of lower than average noise regions
being missed, while conversely, in higher noise regions there is a danger that
spurious noise spikes will be mistakenly interpreted as real sources.

In order to reduce the number of spurious sources and avoid missing
genuine sources we have converted each emission map into a
signal-to-noise ratio map by dividing it by its corresponding noise
map. These maps also contain a large fraction of low surface
brightness diffuse and filamentary emission, particularly towards the
Galactic centre and towards the tangent with the Norma spiral arm and
other prominent star forming regions. This not only complicates the
identification of compact sources but can also lead to the
identification of a large number of spurious sources. To limit the
source confusion due to large scale extended diffuse emission we have
removed it using the program \texttt{FINDBACK}, which is part of the
\texttt{CUPID}\footnote{\texttt{CUPID} is part of the
  \texttt{STARLINK} software suite (http://starlink.jach.hawaii.edu).}
clump identification and analysis package. This produces maps with a
higher contrast between compact sources and their local background. In
Fig.\,\ref{fig:emission_snr_maps} we present an emission map to show
the affect of the varying noise level in the left panel, while in the
right panel we show the same map after smoothing the variations by
dividing through by the noise map and spatial filtering to remove the
diffuse background. These two images nicely illustrate the advantages
of using the signal-to-noise map for the source detection.

To summarise, we first smoothed the noise variations by dividing the maps by
the noise map and created a signal-to-noise map, and as a second step we
applied spatial filtering to remove part of the extended emission. This is
compared to the original maps with varying noise level in
Fig.\,\ref{fig:emission_snr_maps}.

\begin{figure*}
\begin{center}
\includegraphics[width=0.95\textwidth, trim= 0 0 0 0]{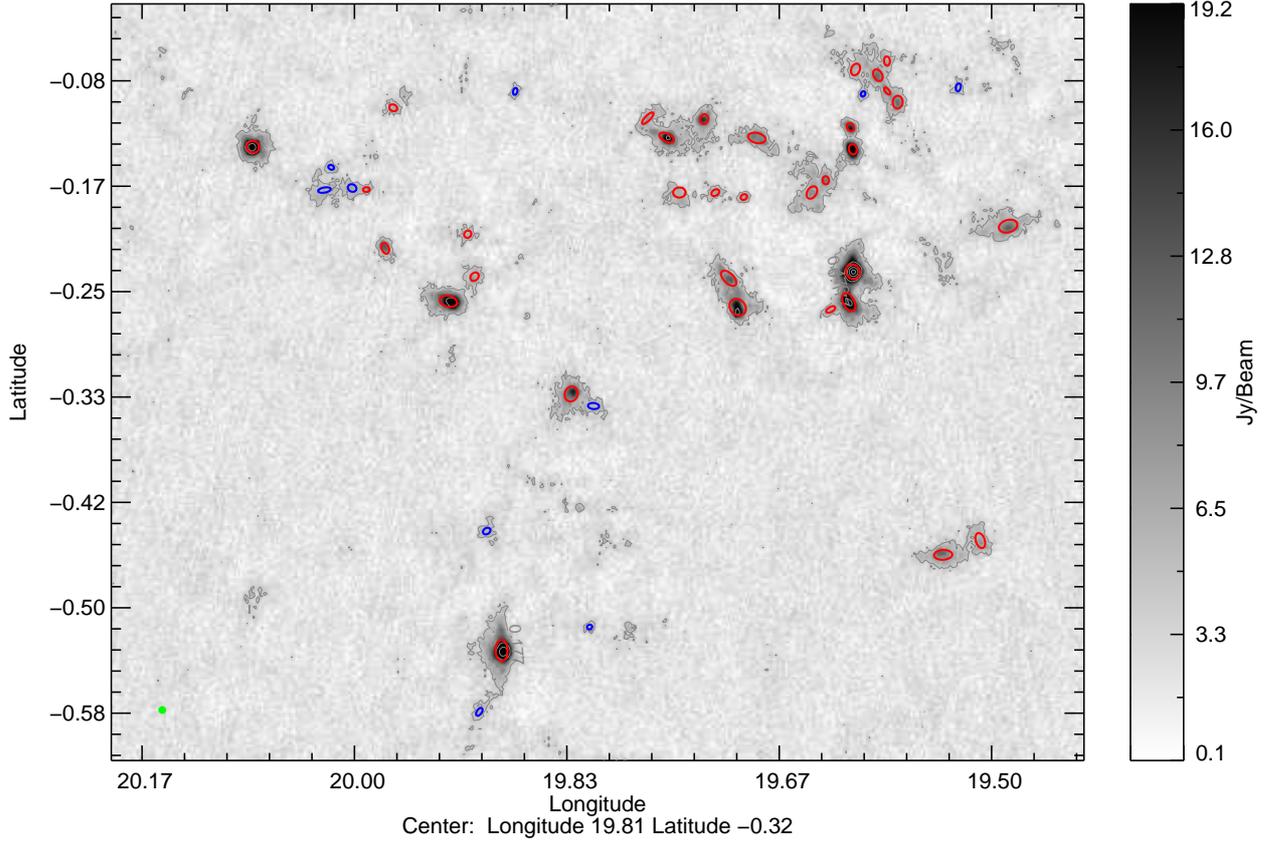}

\caption{\label{fig:example_detection} Example of the sources
  extracted. The field shown in greyscale is the region outlined in
  red in the left panel of Fig.\,\ref{fig:emission_snr_maps}. The
    positions and sizes of sources identified by \sex\ are shown as
    red and blue ellipses.  The red ellipses indicate to the most
    reliable detections, while the blue ellipses indicated the
    detections associated with higher flag values, given by \sex,
    which correspond to less reliable detections (see discussion in
  Sect.\,\ref{cat_description} for details). The ATLASGAL beam size is
  shown in green in the lower left corner of the map. Contours start
  at 3$\sigma$ (i.e., 0.18\,Jy\,beam$^{-1}$) and increase with steps
  of 0.48, 1.61, 4.32, 8.96, 15.89\,Jy\,beam$^{-1}$.}

\end{center}
\end{figure*}

\sex\ can work on two images simultaneously, one for the source detection and
one for measuring the physical parameters of each source. We have used the
spatially filtered signal-to-noise map for source detection, however, we have
used the original emission maps to extract the corresponding source
parameters. In this way, the flux and size of the sources are not affected by
the filtering process performed for the source detection. These steps should
ensure that only genuine sources make it into the final catalogue. In
Fig.\,\ref{fig:example_detection} we present a subregion, taken from the
emission map presented in the left panel of Fig.\,\ref{fig:emission_snr_maps},
showing the positions and sizes of sources identified by \sex.

The ATLASGAL maps were searched for emission above a threshold of
3$\sigma_{\rm{\rms}}$, where $\sigma_{\rm{\rms}}$ is the local noise level
determined from a user defined box by \sex. The value of this threshold
was determined empirically by varying its value for a number of test regions
so as to optimise the number of genuine sources detected while avoiding the
inclusion of the more dubious sources. In addition to the threshold we also
reject any sources with fewer pixels than the beam integral (i.e., $\sim$11 pixels).

\begin{figure}[ht]
\centering \includegraphics[trim = 15mm 4mm 5mm 1mm,clip,width=0.4\textwidth]{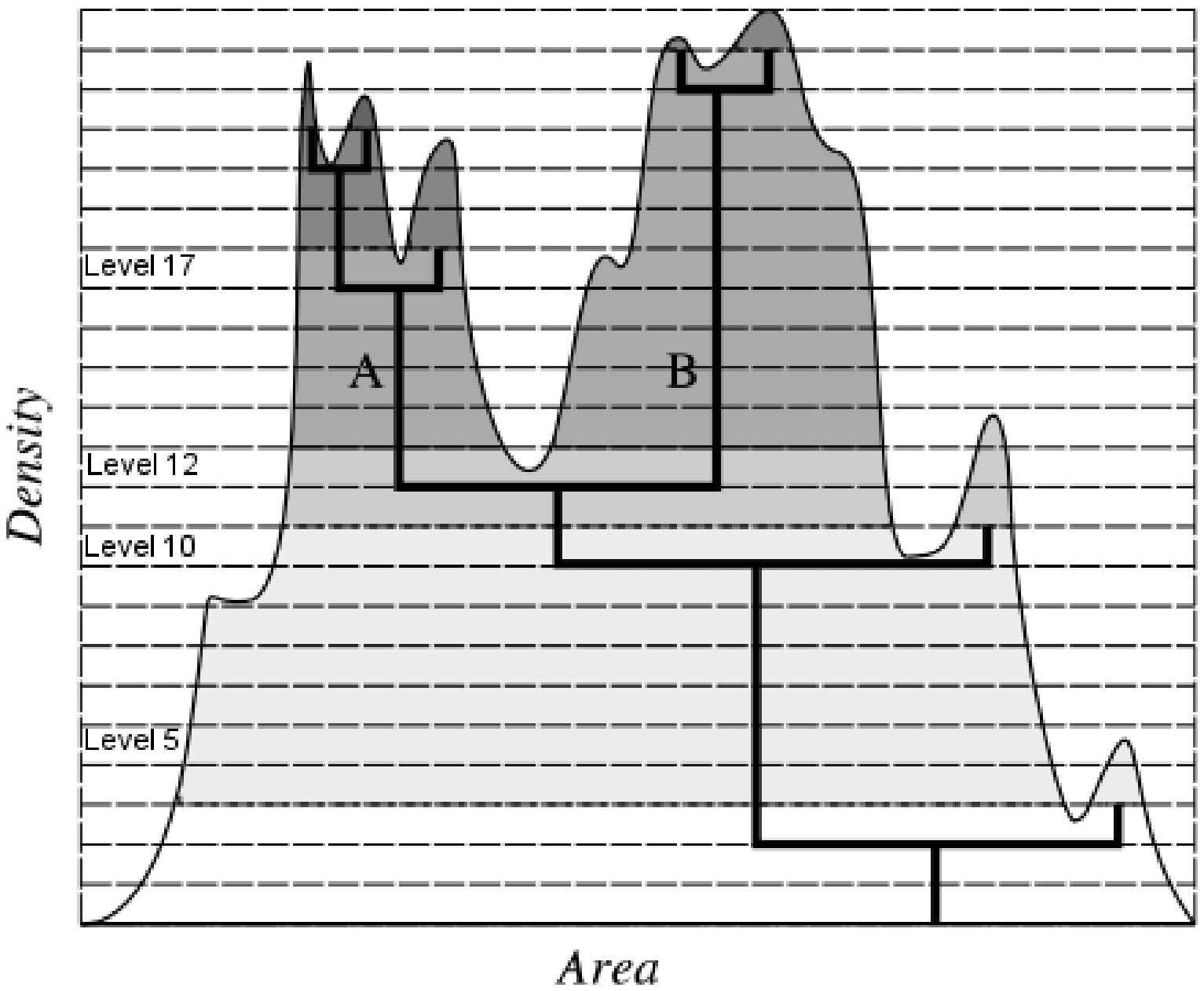}
\caption{Example of the deblending process. The dependence of a pixel to one
  source or another will depend on the relative integrated intensities (grey
  area) of their adjacent pixel for each level (dotted lines). This figure has
  been reproduced here with permission from
  \citet{bertin}.}\label{fig:deblending}
\end{figure}

To separate possible blended sources, \sex\ divided the emission into 40
sub-levels of flux. In each of these 40 levels, the ratio of the integrated
intensities of adjacent pixels is computed.  If the ratio between two pixels
is greater than 0.001, then those pixels are considered part of a different
source. If not, both pixels are considered to be part of the same source. This
comparison is done for each pixel in each level. To illustrate this process we
present a schematic diagram in Fig.\,\ref{fig:deblending} that shows a complex
emission profile that is divided in 23 sub-levels, at the 10th level the
source is separated in two branches then at the 12th level the first branch is
separated again in two sources A and B, at the level 17th there is a small
separation at the branch B, but the ratio of the integrated intensity of this
pseudo new branch is lower than the ratio defined to separate sources so it is
declared as part of the B branch. The process continues until the last level
at which point seven discrete sources have been identified.

\section{Artificial sources}
\label{art}

In order to check the reliability of the compact source catalogue, we test our
extraction algorithm on maps with artificial sources with known properties. To
test the influence of the background emission in the extraction of sources, we
created maps with uniform and varying background noise. These two cases allow
us to compare the performance of the extraction algorithm in the idealised
case where the background is smooth and the more realistic case when the
sources are intermixed with large scale diffuse emission.

The varying background noise map was created from the residual maps produced
by \sex. The field used in this test is the map located at $19.5\degr$, which
is fairly representative of the rest of the ATLASGAL maps. The residual map
contains background emission and zeros at the position of the extracted
sources.  The test map was smoothened using a boxcar function with a width of
10 pixels to remove the sharp ridges left by the source removal. These steps
produced an emission map with varying noise properties similar to the emission
seen in the ATLASGAL maps.

We used the \texttt{CUPID} task \texttt{MAKECLUMPS} to generate a catalogue of
artificial objects with a uniform distribution in $\ell$, $b$. The flux of the
injected objects was uniformly distributed between 0.01 and
0.75\,Jy\,beam$^{-1}$, however, the source size was set to a FWHM of 5 pixels
(30\arcsec) with a Gaussian shape. Taking these input parameters and using the
test field as a template, \texttt{MAKECLUMPS} allowed us to produce a field of
artificial sources with some added noise that has the same dimensions and
coordinates of the test field. To produce the varying background test field we
did not add additional noise with the algorithm but simply co-added the
artificial source map with the varying background noise map described in the
previous paragraph. For the uniform background test field we simply used the
output produced by the \texttt{MAKECLUMPS} task but this time set the noise
level to be similar to the original map (i.e., $\sim$60\,mJy\,beam$^{-1}$).

We repeated this process to produce 20 artificial source maps, all with both a
varying and uniform background noise level. This group of 20 maps produced a
catalogue of $\sim$5000 artificial sources, which is comparable to the number
of genuine sources detected in the region presented here. For the varying
background noise maps we produced background subtracted signal-to-noise maps
following the same procedure used for the real ATLASGAL maps (see
Sect.\,\ref{sex} for further details). \sex\ was then run on these artificial
source maps using the same parameters as those used for the construction of
the real catalogue. To compare the dependence of completeness on the
background, the same extraction algorithm was tested on the maps with uniform
noise. However, for these there was no need to try and filter the large scale
variations in the background and therefore \sex\ was used directly on the
simulated maps.

\begin{figure}[t!]
  \centering
  \includegraphics[width=\linewidth]{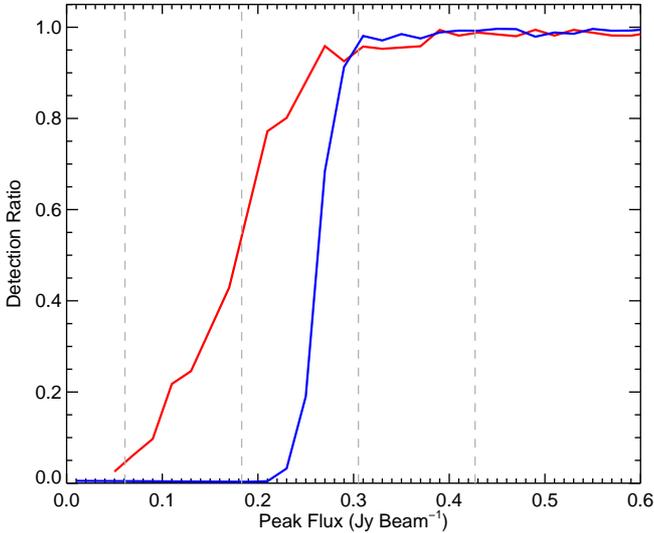}
  \caption{\label{fig:completeness} Detection ratio of recovered to injected
    sources as a function of input peak fluxes. The blue solid and red dashed lines show
    the detection ratios for sources injected onto a uniform and varying
    background noise map, respectively. The dashed vertical lines indicate
    (from left to right) the 1$\sigma$, 3$\sigma$, 5$\sigma$, and 7$\sigma$
    noise levels.  }
\end{figure}

In Fig.\,\ref{fig:completeness} we plot the number of sources
recovered as a function of input peak flux for the uniform and varying
background noise maps (indicated by the blue and red lines,
respectively). In the case of the uniform noise level the source
extraction algorithm is more than 99\% complete above
$\sim$\,0.3\,mJy\,beam$^{-1}$, which corresponds to approximately
5$\sigma$. The shape of the detection ratio and the 5$\sigma$
completeness level found for the uniform background noise maps is very
similar to that found for the BGPS
(\citealt{rosolowsky2010}). However, the completeness profile
determined from the sources injected and recovered from the varying
background noise, which is much more realistic, is very different. In
the case of the uniform background noise maps no sources are detected
below the $\sim$\,3$\sigma$ level, however, we start to detect sources
at $\sim$\,1$\sigma$ when adopting varying-background-noise maps. The
detection of low peak flux sources is better in the varying background
noise maps compared to those with uniform noise. This is because the
peak fluxes are boosted by the diffuse continuum emission they have
been injected onto. As a consequence, the fluxes of weaker sources
($<$5$\sigma$) are somewhat less reliable in this case.

We also note that for the varying background noise case the $\sim$\,99\%
completeness is obtained for peak fluxes above $\sim$\,0.4\,mJy\,beam$^{-1}$,
which, at approximately 6$\sigma$, is slightly worse than that found for the
uniform background noise case. This difference can be attributed to the
presence of the diffuse low surface brightness emission which makes it more
difficult to recover sources below $<$6$\sigma$. Clearly the presence of a
varying background due to large scale diffuse emission has a significant
impact on the detectability and reliability of fluxes for weaker sources
(i.e., $<$6$\sigma$) and therefore we do not advise to use the sources below
6$\sigma$ blindly.

\begin{figure}[t]
\centering
\includegraphics[width=\linewidth]{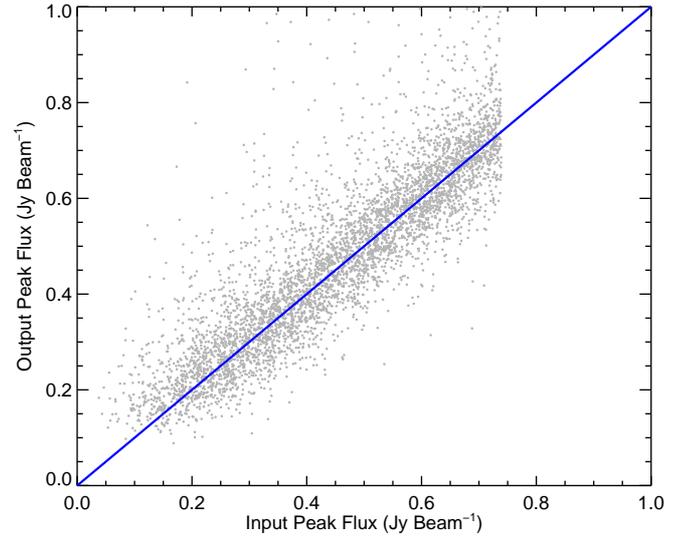}
\includegraphics[width=\linewidth]{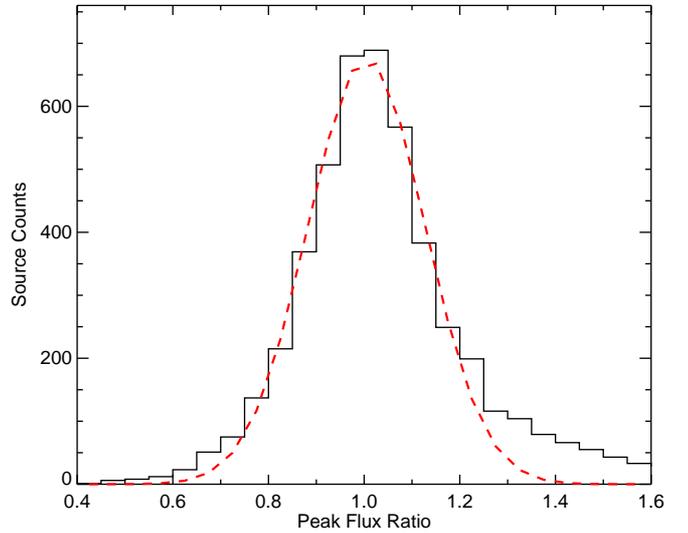}

\caption{\label{fig:diffflux} In the upper panel we present a
  comparison of the recovered values for the peak fluxes values to the
  input values of the injected sources on the maps with varying
  background. The blue line represents equality. In the lower panel we
  present the recovered to injected flux ratio (black histogram) and
  the results of a Gaussian fit to values less than 1.2 (dashed red
  line).}
\end{figure}

In the upper panel of Fig.\,\ref{fig:diffflux} we 
compare peak flux of the recovered and injected sources. It is clear that
these flux values are very well correlated and the injected and recovered
fluxes are in excellent agreement with each other (the blue line plotted over
the data shows indicates equality). We also note the presence of a number of
sources that have significantly higher recovered fluxes than input fluxes,
which is a result of flux boosting from the varying background mentioned
earlier. In the lower panel of Fig.\,\ref{fig:diffflux} we present a histogram
of the ratio of the injected and extracted flux values for each source. The
distribution profile below a ratio value of 1.2 can be approximated by a
Gaussian fit to the data (shown as a dashed red line on the plot) with a peak
at $\sim$\,1 and has a $\sigma$ of 0.12, which would suggests that the flux
values extracted for the sources in this catalogue are generally well
determined with an uncertainty of 12\%, however, for a small number of sources
in more complex regions the flux can be overestimated by up to a factor of 2.

\section{Compact source catalogue}
\label{cat}
\subsection{Catalogue description}
\label{cat_description}
\setcounter{table}{0}

\setlength{\tabcolsep}{5pt}

\begin{table*}

  \begin{comment}

select replace(replace(atlas_name_original,'-','$-$'),'G','AGAL'),'\&',round(gal_long_peak,3),'\&', round(gal_lat_peak,3),'\&',round(gal_long,3),'\&', round(gal_lat,3),'\&', round(major*3600,0),'\&', round(minor*3600,0),'\&',round( 90+pa,0),'\&',if(radius_deconvolved > 0.1,round(2.4*radius_deconvolved,0),'$\\cdots$'),'\&', round(flux_max,2),'\&', round(power(power(image_rms,2)+power(flux_max*0.15,2),0.5),2),'\&',round(flux_auto/11.6,2),'\&',round(power(power(flux_auto_err/(11.6),2)+power((flux_auto/(11.6))*0.15,2),0.5),2),'\&',flags,'\&',round(flux_max/image_rms,1),'\\\\' from atlas_extractor_3_2sigma where (gal_long between 19.41 and 20.1666 and gal_lat between -0.6 and -0.03) and isoarea_image > 10 and flux_auto > 0 and flux_max > 0 and reject_flag is null and image_name is not null order by atlas_name_original

\end{comment}

\begin{center}\caption{\label{tbl:cattable} The ATLASGAL compact source catalogue.  The columns are as follows: (1) name derived from Galactic coordinates of the maximum intensity in the source; (2)-(3) Galactic coordinates of maximum intensity in the catalogue source; (4)-(5) Galactic coordinates of emission centroid; (6)-(8) semi-major and semi-minor size and source position angle measured anti-clockwise from Galactic north; (9) effective radius of source; (10)-(13) peak and integrated flux densities and their associated uncertainties; (14) \sex\ detection flag (see text for details); (15) signal to noise ratio (SNR) --- values for sources with peak flux below $6\sigma$ detection should not be used blindly.}
\begin{minipage}{\linewidth}
%\scriptsize
\begin{tabular}{l....rrrr....c.}
  \hline \hline
  \multicolumn{1}{c}{Name} 
  &  \multicolumn{1}{c}{$\ell_{\mathrm{max}}$} &  \multicolumn{1}{c}{$b_{\mathrm{max}}$}
  &  \multicolumn{1}{c}{$\ell$} &  \multicolumn{1}{c}{$b$} &
  \multicolumn{1}{c}{$\sigma_{\rm{maj}}$} &  \multicolumn{1}{c}{$\sigma_{\rm{min}}$} &  \multicolumn{1}{c}{PA} &
  \multicolumn{1}{c}{$\theta_{\rm{R}}$} &  \multicolumn{1}{c}{$S_{\rm{peak}}$} & \multicolumn{1}{c}{$\Delta S_{\rm{peak}}$} & \multicolumn{1}{c}{$S_{\rm{int}}$}& \multicolumn{1}{c}{$\Delta S_{\rm{int}}$} & \multicolumn{1}{c}{Flag} & \multicolumn{1}{c}{SNR} \\
   
  \multicolumn{1}{c}{} &  \multicolumn{1}{c}{($^{\circ}$)} &
  \multicolumn{1}{c}{($^{\circ}$)} &  \multicolumn{1}{c}{($^{\circ}$)} &
  \multicolumn{1}{c}{($^{\circ}$)} &  \multicolumn{1}{c}{($''$)}
  &  \multicolumn{1}{c}{($''$)} &  \multicolumn{1}{c}{($^{\circ}$)}&  \multicolumn{1}{c}{($''$)}
  &  \multicolumn{2}{c}{(Jy\,beam$^{-1}$)} &  \multicolumn{2}{c}{(Jy)}&\\
 
  \multicolumn{1}{c}{(1)} &  \multicolumn{1}{c}{(2)} &  \multicolumn{1}{c}{(3)} &  \multicolumn{1}{c}{(4)} &
  \multicolumn{1}{c}{(5)} &  \multicolumn{1}{c}{(6)} &  \multicolumn{1}{c}{(7)} &  \multicolumn{1}{c}{(8)} &
  \multicolumn{1}{c}{(9)} &  \multicolumn{1}{c}{(10)} &  \multicolumn{1}{c}{(11)} & \multicolumn{1}{c}{(12)} &  \multicolumn{1}{c}{(13)} &  \multicolumn{1}{c}{(14)} &  \multicolumn{1}{c}{(15)} \\
  \hline
AGAL019.486$-$00.199	&	19.486	&	-0.199	&	19.487	&	-0.198	&	27	&	17	&	73	&	48	&	0.80	&	0.13	&	9.35	&	1.67	&	2	&	12.9	\\
AGAL019.508$-$00.447	&	19.508	&	-0.447	&	19.509	&	-0.447	&	22	&	13	&	161	&	34	&	0.48	&	0.10	&	4.06	&	0.86	&	0	&	7.6	\\
AGAL019.526$-$00.086	&	19.526	&	-0.086	&	19.526	&	-0.089	&	11	&	7	&	14	&	$\cdots$	&	0.35	&	0.08	&	1.59	&	0.45	&	0	&	5.6	\\
AGAL019.539$-$00.456	&	19.539	&	-0.456	&	19.538	&	-0.458	&	26	&	14	&	87	&	40	&	0.76	&	0.13	&	7.72	&	1.42	&	2	&	11.2	\\
AGAL019.572$-$00.101	&	19.572	&	-0.101	&	19.574	&	-0.100	&	18	&	14	&	4	&	33	&	0.71	&	0.13	&	4.13	&	0.87	&	2	&	8.7	\\
AGAL019.581$-$00.069	&	19.581	&	-0.069	&	19.582	&	-0.068	&	13	&	8	&	176	&	11	&	0.48	&	0.10	&	3.90	&	0.83	&	3	&	6.5	\\
AGAL019.582$-$00.091	&	19.582	&	-0.091	&	19.582	&	-0.091	&	12	&	6	&	142	&	$\cdots$	&	0.59	&	0.13	&	1.81	&	0.49	&	3	&	6.5	\\
AGAL019.589$-$00.079	&	19.589	&	-0.079	&	19.589	&	-0.079	&	18	&	12	&	150	&	28	&	0.73	&	0.14	&	9.27	&	1.66	&	3	&	8.3	\\
AGAL019.601$-$00.092	&	19.601	&	-0.092	&	19.601	&	-0.094	&	8	&	6	&	24	&	$\cdots$	&	0.40	&	0.11	&	1.49	&	0.43	&	0	&	4.2	\\
AGAL019.606$-$00.072	&	19.606	&	-0.072	&	19.607	&	-0.074	&	17	&	12	&	22	&	28	&	0.46	&	0.11	&	6.35	&	1.22	&	3	&	5.6	\\
AGAL019.609$-$00.137	&	19.609	&	-0.137	&	19.609	&	-0.138	&	15	&	11	&	169	&	23	&	2.79	&	0.43	&	7.59	&	1.40	&	3	&	32.3	\\
AGAL019.609$-$00.234	&	19.609	&	-0.234	&	19.609	&	-0.234	&	22	&	19	&	151	&	45	&	19.20	&	2.88	&	44.85	&	7.01	&	3	&	252.3	\\
AGAL019.611$-$00.121	&	19.611	&	-0.121	&	19.611	&	-0.120	&	13	&	10	&	147	&	18	&	1.17	&	0.20	&	2.86	&	0.66	&	3	&	13.8	\\
AGAL019.612$-$00.259	&	19.612	&	-0.259	&	19.612	&	-0.258	&	28	&	16	&	153	&	46	&	1.88	&	0.29	&	14.88	&	2.51	&	3	&	25.7	\\
AGAL019.628$-$00.264	&	19.628	&	-0.264	&	19.626	&	-0.264	&	14	&	7	&	60	&	$\cdots$	&	0.46	&	0.10	&	1.62	&	0.45	&	3	&	6.8	\\
AGAL019.631$-$00.162	&	19.631	&	-0.162	&	19.630	&	-0.162	&	12	&	9	&	175	&	13	&	0.70	&	0.14	&	5.80	&	1.13	&	3	&	7.9	\\
AGAL019.641$-$00.172	&	19.641	&	-0.172	&	19.641	&	-0.172	&	19	&	13	&	32	&	32	&	0.53	&	0.12	&	11.06	&	1.93	&	2	&	5.7	\\
AGAL019.686$-$00.127	&	19.686	&	-0.127	&	19.684	&	-0.129	&	25	&	14	&	103	&	40	&	0.70	&	0.14	&	7.52	&	1.39	&	2	&	8.2	\\
AGAL019.694$-$00.174	&	19.694	&	-0.174	&	19.694	&	-0.175	&	10	&	8	&	62	&	$\cdots$	&	0.44	&	0.10	&	1.02	&	0.34	&	0	&	6.0	\\
AGAL019.699$-$00.266	&	19.699	&	-0.266	&	19.699	&	-0.262	&	26	&	21	&	138	&	52	&	1.97	&	0.30	&	12.69	&	2.18	&	3	&	27.7	\\
AGAL019.706$-$00.239	&	19.706	&	-0.239	&	19.706	&	-0.239	&	27	&	14	&	136	&	41	&	0.88	&	0.15	&	9.56	&	1.71	&	3	&	12.1	\\
AGAL019.716$-$00.171	&	19.716	&	-0.171	&	19.717	&	-0.172	&	12	&	8	&	58	&	10	&	0.44	&	0.10	&	2.62	&	0.63	&	0	&	5.7	\\
AGAL019.726$-$00.114	&	19.726	&	-0.114	&	19.726	&	-0.114	&	14	&	11	&	10	&	23	&	1.09	&	0.18	&	7.61	&	1.41	&	0	&	12.5	\\
AGAL019.746$-$00.171	&	19.746	&	-0.171	&	19.745	&	-0.172	&	18	&	14	&	88	&	33	&	0.40	&	0.09	&	3.93	&	0.84	&	2	&	6.0	\\
AGAL019.754$-$00.129	&	19.754	&	-0.129	&	19.755	&	-0.128	&	22	&	11	&	116	&	30	&	1.87	&	0.29	&	11.23	&	1.96	&	2	&	20.9	\\
AGAL019.771$-$00.114	&	19.771	&	-0.114	&	19.770	&	-0.113	&	22	&	7	&	45	&	$\cdots$	&	0.54	&	0.10	&	2.78	&	0.65	&	2	&	8.3	\\
AGAL019.811$-$00.342	&	19.811	&	-0.342	&	19.812	&	-0.340	&	16	&	9	&	95	&	15	&	0.37	&	0.08	&	3.27	&	0.73	&	19	&	6.0	\\
AGAL019.817$-$00.516	&	19.817	&	-0.516	&	19.816	&	-0.515	&	8	&	6	&	68	&	$\cdots$	&	0.33	&	0.08	&	1.20	&	0.38	&	0	&	5.3	\\
AGAL019.829$-$00.329	&	19.829	&	-0.329	&	19.830	&	-0.331	&	22	&	18	&	20	&	43	&	1.15	&	0.19	&	9.90	&	1.76	&	2	&	17.2	\\
AGAL019.874$-$00.092	&	19.874	&	-0.092	&	19.874	&	-0.092	&	10	&	6	&	13	&	$\cdots$	&	0.32	&	0.08	&	1.08	&	0.36	&	0	&	5.3	\\
AGAL019.882$-$00.534	&	19.882	&	-0.534	&	19.885	&	-0.534	&	30	&	17	&	178	&	50	&	7.43	&	1.12	&	27.37	&	4.39	&	2	&	90.7	\\
AGAL019.899$-$00.441	&	19.899	&	-0.441	&	19.896	&	-0.439	&	11	&	9	&	56	&	10	&	0.35	&	0.08	&	2.51	&	0.61	&	18	&	5.7	\\
AGAL019.902$-$00.582	&	19.902	&	-0.582	&	19.902	&	-0.582	&	13	&	7	&	35	&	$\cdots$	&	0.38	&	0.09	&	3.24	&	0.73	&	16	&	5.9	\\
AGAL019.906$-$00.241	&	19.906	&	-0.241	&	19.906	&	-0.238	&	13	&	10	&	44	&	20	&	0.45	&	0.10	&	3.33	&	0.74	&	0	&	5.8	\\
AGAL019.911$-$00.206	&	19.911	&	-0.206	&	19.911	&	-0.205	&	11	&	10	&	45	&	15	&	0.33	&	0.08	&	2.30	&	0.57	&	0	&	4.9	\\
AGAL019.922$-$00.259	&	19.922	&	-0.259	&	19.926	&	-0.258	&	25	&	15	&	103	&	42	&	2.51	&	0.38	&	13.96	&	2.37	&	0	&	36.8	\\
AGAL019.967$-$00.106	&	19.967	&	-0.106	&	19.970	&	-0.105	&	11	&	9	&	116	&	14	&	0.45	&	0.09	&	2.29	&	0.57	&	0	&	7.2	\\
AGAL019.977$-$00.214	&	19.977	&	-0.214	&	19.976	&	-0.216	&	16	&	10	&	158	&	21	&	0.87	&	0.15	&	3.12	&	0.70	&	0	&	13.9	\\
AGAL019.991$-$00.169	&	19.991	&	-0.169	&	19.991	&	-0.169	&	10	&	7	&	86	&	$\cdots$	&	0.55	&	0.11	&	1.52	&	0.43	&	3	&	7.9	\\
AGAL020.004$-$00.167	&	20.004	&	-0.167	&	20.002	&	-0.168	&	13	&	10	&	115	&	19	&	0.40	&	0.09	&	4.63	&	0.95	&	18	&	5.7	\\
AGAL020.019$-$00.151	&	20.019	&	-0.151	&	20.018	&	-0.152	&	9	&	6	&	119	&	$\cdots$	&	0.37	&	0.09	&	1.96	&	0.51	&	16	&	5.0	\\
AGAL020.022$-$00.169	&	20.022	&	-0.169	&	20.024	&	-0.170	&	18	&	7	&	80	&	$\cdots$	&	0.38	&	0.09	&	5.80	&	1.13	&	19	&	5.5	\\
AGAL020.081$-$00.136	&	20.081	&	-0.136	&	20.080	&	-0.136	&	18	&	16	&	106	&	36	&	7.68	&	1.15	&	18.43	&	3.04	&	0	&	119.8	\\
  \hline
\end{tabular}\\
\end{minipage}
Notes: Only a small portion of the data is provided here, the full table is only  available in electronic form at the CDS via anonymous ftp to cdsarc.u-strasbg.fr (130.79.125.5) or via http://cdsweb.u-strasbg.fr/cgi-bin/qcat?J/A+A/.
\end{center}
\end{table*}

\setlength{\tabcolsep}{6pt}

\sex\ was run on each of the seventeen $3\degr \times 3\degr$ fields that make
up the inner most portion of the ATLASGAL survey (i.e., $330\degr < \ell <
21\degr$). The results obtained for each field were subsequently combined in
order to compile a complete catalogue of sources, taking care to eliminate
duplicated sources located in the overlap regions between adjoining fields. In
total we have identified \nsources\ sources in the 153 deg$^2$.

In Table\,\ref{tbl:cattable} we present a sample of the catalogue (for
the region presented in Fig.\,\ref{fig:example_detection}) as an
example of its form and content. The full and up-to-date catalogue and
ATLASGAL emission maps will be hosted by the ATLASGAL project
page\footnote{\url{www.mpifr-bonn.mpg.de/div/atlasgal/}}, the
ESO\footnote{Available through \url{http://archive.eso.org/wdb/wdb/adp/phase3_main/form}} and CDS\footnote{\url{http://cdsweb.u-strasbg.fr/cgi-bin/qcat?J/A+A/}}
database. There will be incremental improvements in image processing,
calibration and source extraction. In the table we present the source
name, the source peak and barycentric position, the size of the
semi-major and minor axis and their position angle, the deconvolved
radius, the peak and integrated fluxes and their associated errors and
any warning flags generated by the algorithm (see full bit per bit
description below).

The source names are based on the Galactic coordinates of the peak flux
position, which are given in Cols.\,2 and 3. The barycentric coordinates are
determined from the first order moments of the longitude and latitude
profiles. The barycentric position generally defines the centre position of a
source, however, if the emission profile of a source is strongly skewed or is
associated with substructures, the peak position can be a more accurate
measure of a source's position. The source sizes describe the detected source
as an elliptical shape, although no fitting is actually performed. The
semi-major and semi-minor axis lengths represent the standard deviation of the
pixel co-ordinate values about the centroid position, weighted by the pixel
values. The position angle is given as anti-clockwise from Galactic north. 

Since the source sizes are determined only using the pixels above the
detection threshold it is likely these underestimate the true source
sizes, and in some cases results in source sizes that are smaller than
the beam. Following \citet{rosolowsky2010} we estimate the angular
radius ($\theta_{\rm{R}}$) from the geometric mean of the deconvolved
major and minor axes and multiplied by a factor $\eta$ that relates
the \rms\ size of the emission distribution of the source to its
angular radius (Eqn.\,6 of \citealt{rosolowsky2010}):

\begin{equation}
\label{radius}
\theta_{\rm{R}}= \eta \left[(\sigma_{\rm{maj}}^2-\sigma_{\rm{bm}}^2)(\sigma_{\rm{min}}^2-\sigma_{\rm{bm}}^2)\right]^{1/4},
\end{equation}

\noindent where $\sigma_{bm}$ is the \rms\ size of the beam (i.e.,
$\sigma_{bm}=\theta_{\mathrm{FWHM}}/\sqrt{8\ln 2} \simeq 8''$). The value of
$\eta$ depends on the size of the source with respect to the beam and the
emission distribution. \citet{rosolowsky2010} adopted a value of 2.4, which is
the median of the values derived from a range of models, however, they note
that the value of $\eta$ can vary by a factor of 2 in their simulations. For
consistency and to facilitate comparisons between the BGPS and ATLASGAL
catalogues we also adopt this value, which is approximately equivalent to the sources effective radius (i.e., $R_{\rm{eff}}=\sqrt(A/\pi)$, where $A$ is the surface area of the source; \citealt{dunham2011}).

The peak flux is directly obtained from \sex, however, since the algorithm
treats each pixel in the given emission map in a statistically independent
manner and the ATLASGAL maps are gridded with $\sim$3 pixels per beam it is
necessary to account for the beam area to determine the final measured fluxes
and their associated uncertainties. This was simply done by dividing the derived flux
values by the beam area (i.e., $1.133 \times FWHM^2 \simeq 11.6$ pixels). The errors given for the peak and integrated fluxes include the absolute calibration uncertainty of 15\% mentioned in
Sect.\,2 (see also \citealp{Schuller} for details) combined in quadrature with
the intrinsic measurement error. In the case of the integrated flux this is
provided by \sex, however, the error in the peak flux is derived from the
local \rms\ noise in the image.

In Col.\,15 we present the quality flag given by \sex\ which
gathers all the information about possible problems or artefacts
affecting the source \citep{bertin}. For each source, this flag is
either zero (no particular problem), or is equal to the sum of one or
more number(s) with the following meanings: (1) the object has
neighbours, bright and close enough to significantly bias the
photometry, or bad pixels (more than 10\% of the integrated area
affected); (2) the object was originally blended with another one; (4)
at least one pixel of the object is saturated (or very close to); (8)
the object is truncated (too close to an image boundary); (16)
object's aperture data are incomplete or corrupted; (32) object's
isophotal data are incomplete or corrupted; (64) a memory overflow
occurred during deblending; (128) a memory overflow occurred during
extraction. For example, a flag value of 10 means that the object was
originally blended with another source \emph{and} that it is truncated
because is located too close to the edge of the map. In the final
  column of the table we present the signal to noise for each source.

In Fig.\,\ref{fig:example_detection} we overplot the positions of
sources detected by \sex\ towards a small region to demonstrate its
performance. To check the reliability of sources with high flag values
we plot sources with flag values less than 4 (which are the most
reliable) in red while those with higher flag values are shown in blue. In this example region it is clear that the sources with the
higher flag values are associated with genuine emission, however, the
emission does appear to be weaker and more diffuse than found to be
associated with the lower flag value detections. A similar situation
is seen in other regions examined and so we conclude that these high
flag detections do identify genuine sources of emission, but with the
caveat that their associated parameters are somewhat less reliable.

\subsection{Catalogue properties}

In this section we will look at the Galactic distribution of sources detected
and the overall distribution of their derived parameters. As a check on the
reliability of the ATLASGAL catalogue we compare our results to those
obtained from the Bolocam Galactic Plane Survey (BGPS; \citealt{aguirre2011})
that were recently presented by \citet{rosolowsky2010}.

The BGPS covers 150\,deg$^2$ of the Galactic plane, including the majority of
the first quadrant with a latitude range of $|b| < 0.5\degr$. Although the
overlap between the BGPS and the ATLASGAL region presented in this paper is
relatively modest ($\sim$30\,deg$^2$) when compared to the total area covered
by the two surveys, the overlapping region does contain some of the highest
density regions in the plane, and thus, provides a large number of sources in
common to both surveys. Another thing to bear in mind when comparing the two
catalogues is that the two surveys were preformed at different wavelengths and
hence they have different spatial resolution and sensitivity and so we do not
expect a one to one correlation between individual sources or their
parameters. However, we do expect the overall distribution of source
parameters to be correlated. To facilitate the comparisons between the two
catalogues we have followed the structure and present the same parameters in
Table\,1 as are given in the BGPS catalogue (cf. Table\,1;
\citealt{rosolowsky2010}).

\begin{figure*}
\begin{center}
\includegraphics[width=.45\textwidth, trim= 0 0 0 0]{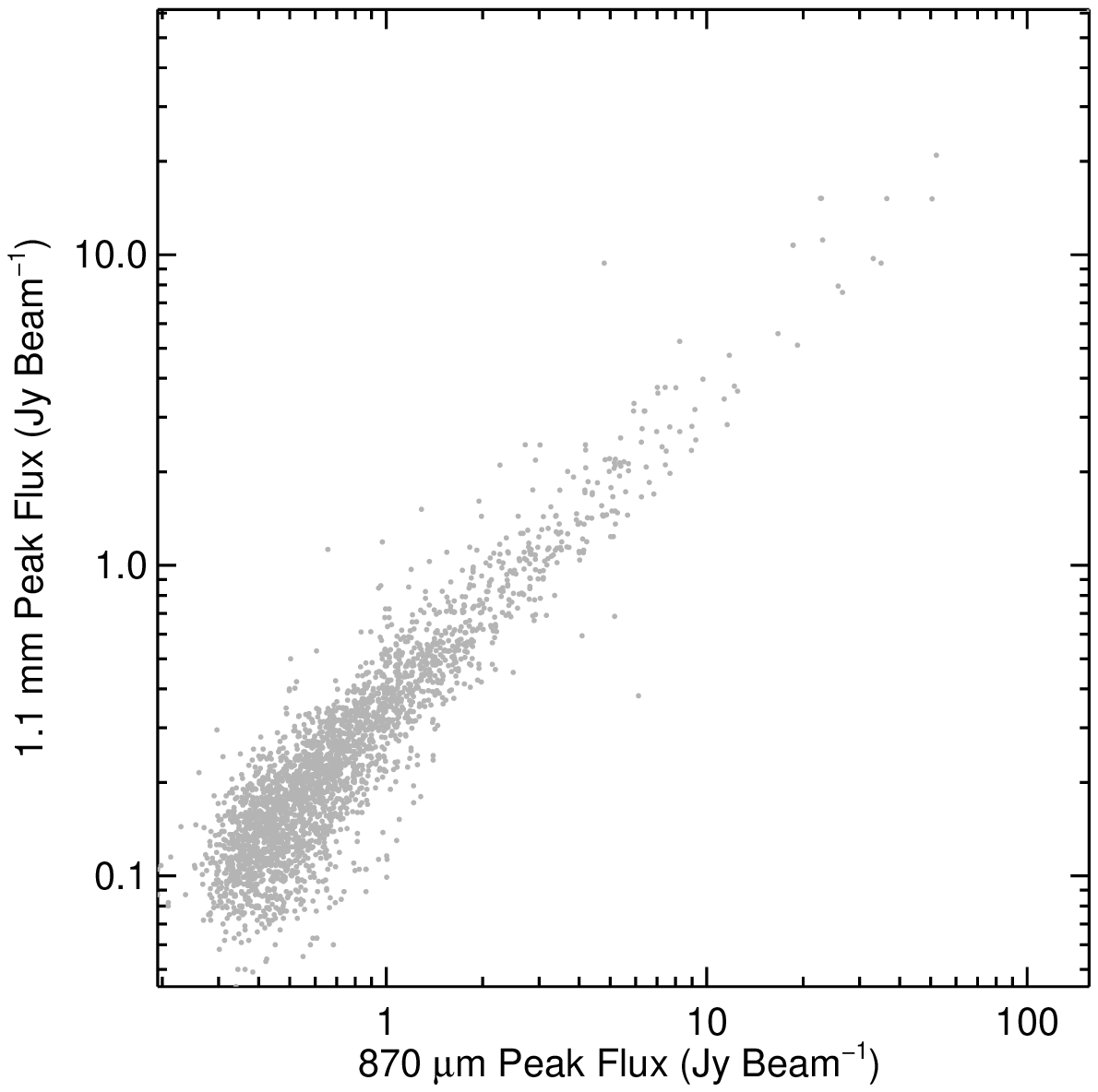}
\includegraphics[width=.45\textwidth, trim= 0 0 0 0]{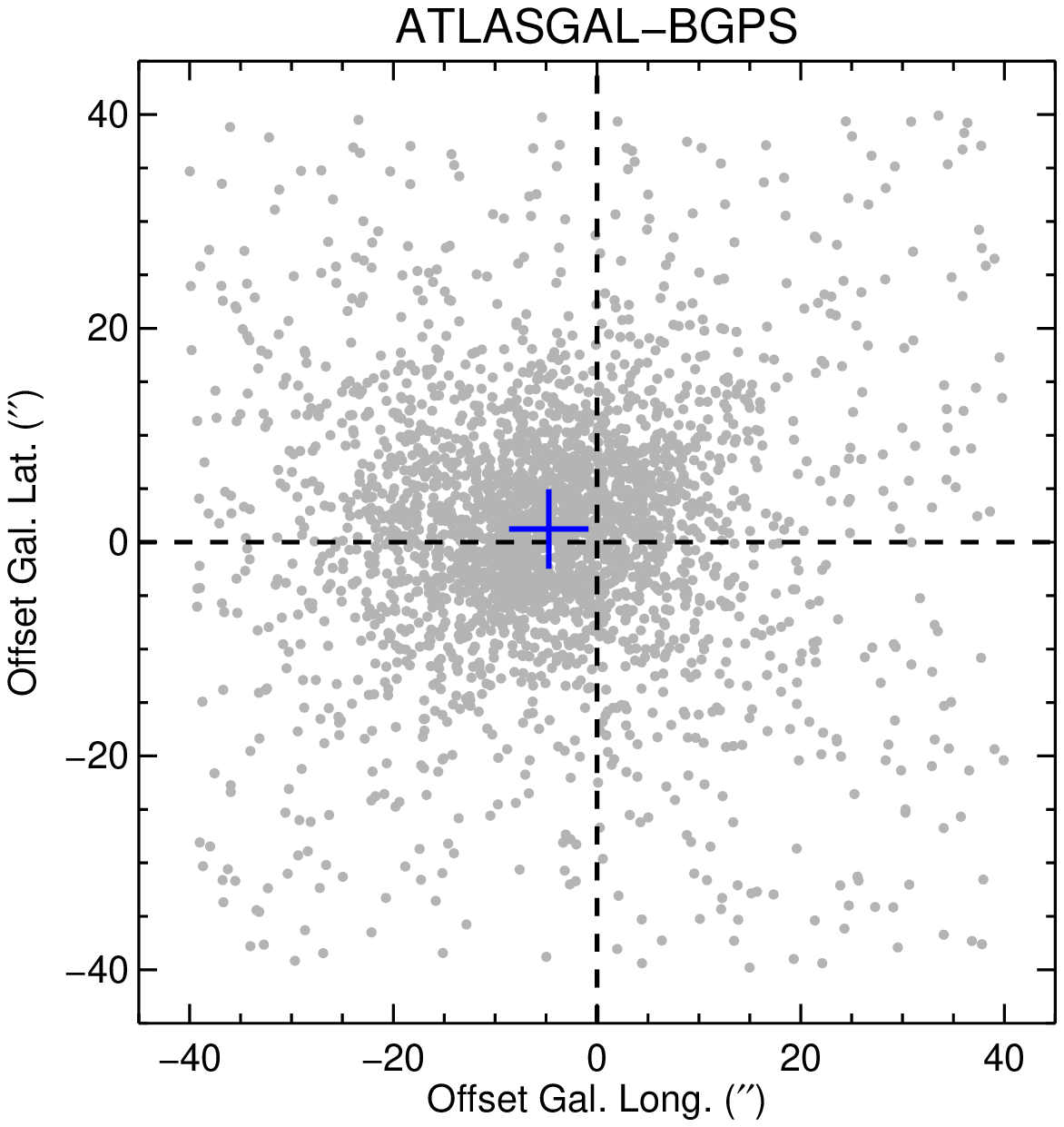}\\
\includegraphics[width=0.45\textwidth, trim= 0 0 0 0]{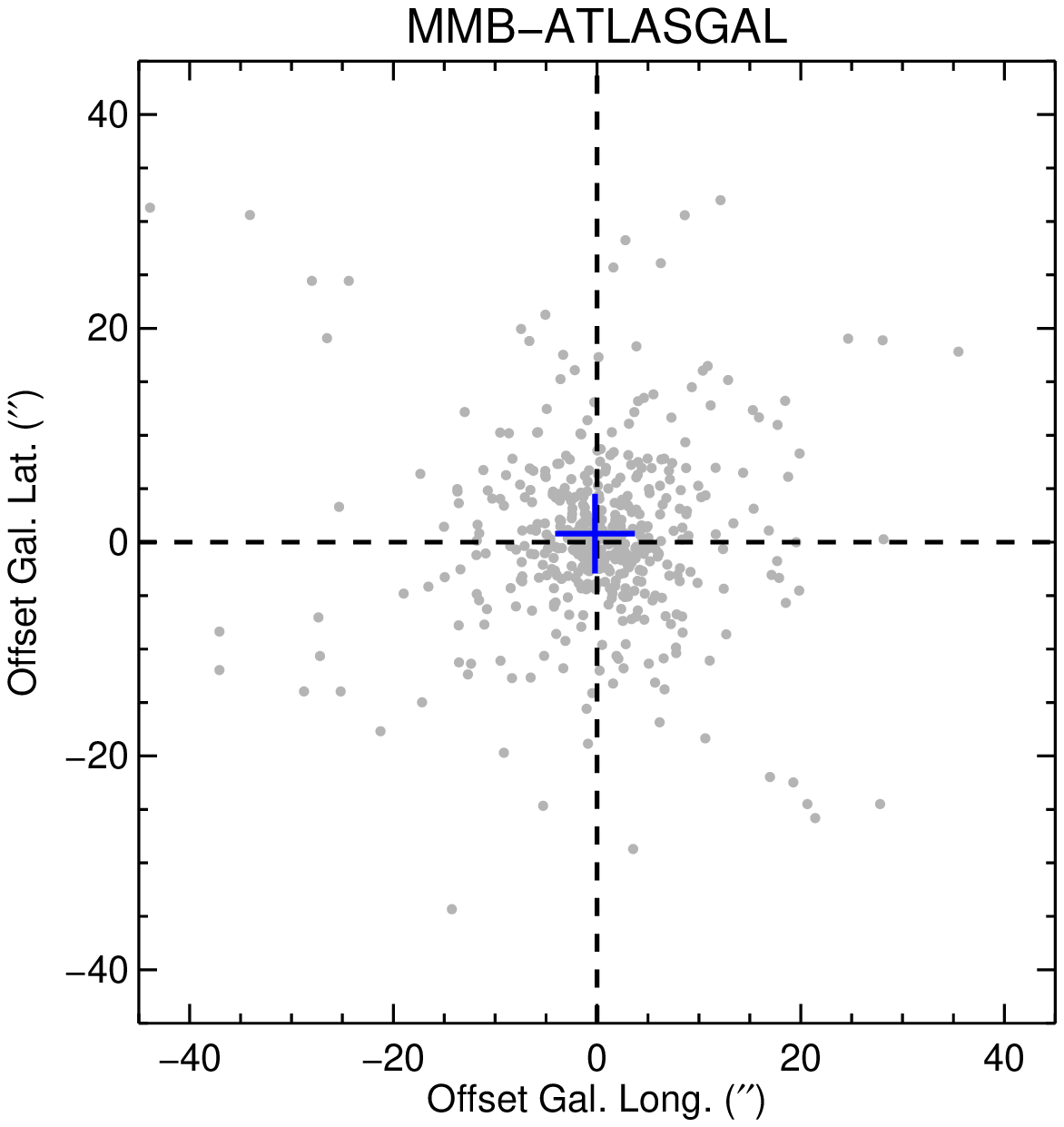}
\includegraphics[width=0.45\textwidth, trim= 0 0 0 0]{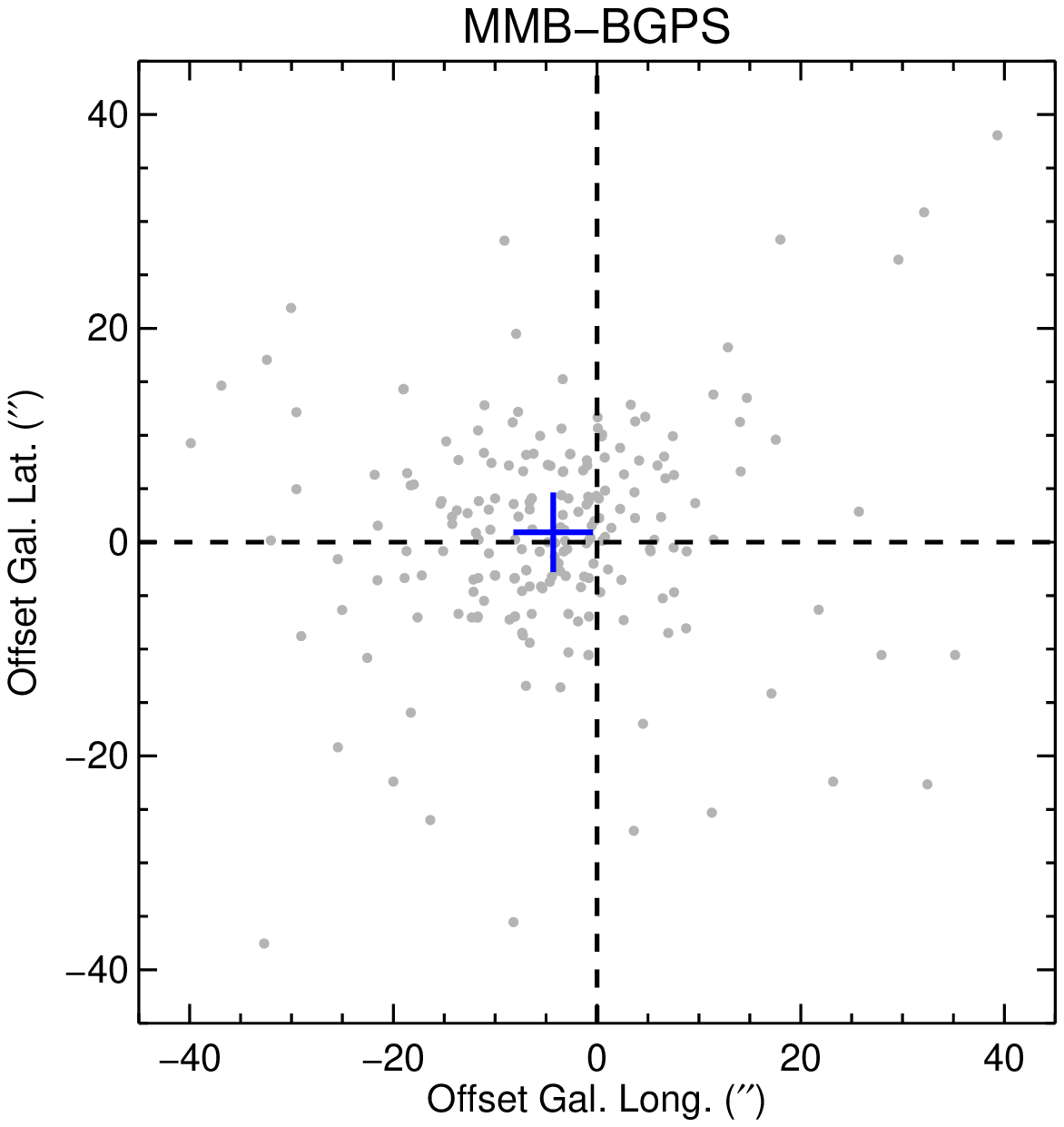}

\caption{\label{fig:astrometry} Results of the cross-matching preformed
  between the ATLASGAL, BGPS and MMB catalogues. In the upper left panel we
  plot the peak fluxes for BGPS and ATLASGAL sources found within a matching
  radius of 40\arcsec. In the upper right panel we compare the angular offsets
  in Galactic longitude and latitude between the ATLASGAL and BGPS
  catalogues. In the two lower panels we present plots showing the angular
  offsets in $\ell$ and $b$ between matched MMB-ATLASGAL (left) and MMB-BGPS
  (right) sources. The dashed lines overplotted in the in the upper right and
  lower panels show the zero offset positions in $\ell$ and $b$, while the
  blue crosses indicate the mean offset of the matched samples.}

\end{center}
\end{figure*}

\subsubsection{Astrometry}

In order to check the astrometry of the ATLASGAL catalogue we compared
the peak positions determined in our catalogue with those determined
by the BGPS.  Using an angular search radius of 40\arcsec\ we identify
$\sim$3,000 likely associations. The matching radius was chosen to be
a little larger than the resolution of the BGPS to obtain a large 
sample of matched sources to compare the positions, while
it is still small enough to avoid multiple matches between sources. In
the upper left panel of Fig.\,\ref{fig:astrometry} we plot the peak
870\,$\mu$m and 1.1\,mm flux densities of the associated sources. As
this plot shows, there is a strong correlation between the fluxes of
the matched sources which suggests that the matched sources are really
associated with each other.

In the upper right panel of Fig.\,\ref{fig:astrometry} we plot the
angular position offsets in Galactic Longitude and Latitude between
the matched ATLASGAL and BGPS sources. Although we should expect a
spread in the angular offsets for individual matched pairs of sources,
given the large number of matches we also expect the mean of the
offset distribution to be centred around zero. However, visual
inspection of this plot reveals a systematic offset between the two
catalogues of $\sim$5\arcsec\ in latitude (see
Table\,\ref{tbl:offsets} for details). This offset is comparable to
the pixel sizes 
for the two surveys (6\arcsec\ and 7\arcsec\ for ATLASGAL and BGPS,
respectively). Given that this offset is significantly smaller than
their respective beam size, and that the BGPS and ATLASGAL sources are
not point sources, means that this systematic offset cannot be seen
from a comparison of contoured emission maps of individual
sources. However, the offset is approximately 20 times the standard
error and is therefore significant.

In an effort to track down the origin of this offset we repeated the matching
procedure described at the beginning of this subsection with the Methanol
Multibeam (MMB) survey catalogue (\citealt{green+2009}). The MMB catalogue
consists of $\sim$700 methanol masers observed at high resolution with the
Australia Telescope Compact Array which has provided positions accurate to
better than an arcsecond \rms\ (\citealt{caswell2010}). Methanol masers are
associated with the early stages of high-mass star formation, which takes
place within deeply embedded dust clumps. We should therefore expect a strong
positional correlation between the methanol masers and dust clumps identified
by ATLASGAL. We have matched $\sim$500 masers with ATLASGAL sources and
present a plot of their positional offsets in the lower left panel of
Fig.\,\ref{fig:astrometry}. It is clear from this plot and from the mean
offsets and standard errors presented in Table\,\ref{tbl:offsets} that there
is no significant offset between the two catalogues.

We also computed the positional offsets between the BGPS and MMB catalogues
(see lower right panel of Fig.\,\ref{fig:astrometry}). We find a very
similar systematic offset ($\Delta \ell \approx $5\arcsec) as found from
the comparison of the ATLASGAL and BGPS catalogues shown in the upper
right panel of Fig.\,\ref{fig:astrometry}. Although the number of
matches found between the MMB and BGPS samples is smaller ($\sim$200)
the mean offset in longitude is five times the standard error and is
therefore significant.

\citet{aguirre2011} checked the astrometry of the BGPS catalogue against the
SCUBA Legacy Catalogue (\citealt{di_francesco2008}; hereafter SLC) and found
the offsets between them to be $\Delta\ell=-$1.8\arcsec$\pm$1.2\arcsec\ and
$\Delta b=0.4$\arcsec$\pm$0.8\arcsec.\footnote{In the \citet{aguirre2011}
  paper they give the offsets determined from the difference between the BGPS
  and SCUBA fields, however, to be consistent with the way the offsets have
  been determined here these values have been reversed.} The mean offset
values and their associated error reported by \citet{aguirre2011} are
consistent with the zero offset position in longitude and latitude. However,
because their standard errors are quite large their results are consistent
with the values determined here for the offset between the ATLASGAL and BGPS
sources (i.e., within 3 standard errors). The excellent angular correlation
between the MMB and ATLASGAL sources, and the fact that the same systematic
offset is observed when comparing the ATLASGAL and MMB catalogues with the
BGPS, would suggest that the observed systematic offset originates from the
BGPS catalogue.

\begin{table}

  \begin{center}\caption{\label{tbl:offsets} Angular offsets between peak flux positions determined from the ATLASGAL, BGPS and MMB catalogues. The errors given are the standard errors.}
\begin{minipage}{\linewidth}
%\scriptsize
\begin{tabular}{lccr}
  \hline \hline
  \multicolumn{1}{l}{Catalogues} &  \multicolumn{1}{c}{$\Delta\ell$}
  &  \multicolumn{1}{c}{$\Delta b$} &  \multicolumn{1}{c}{Matches}\\
  \hline
ATLASGAL-BGPS & $-4.74\pm0.26$ &   $+1.23\pm0.23$ & 2803 \\
MMB-ATLASGAL  & $-0.20\pm0.51$  & $+0.79\pm0.43$ & 459\\
MMB-BGPS      & $-4.30\pm0.86$ & $+0.92\pm0.70$ & 209 \\  
SCUBA-BGPS\footnote{These results have been taken from \citet{aguirre2011}. }      & $-1.8\pm1.2$ & $+0.4\pm0.8$ & $\cdots$ \\  

\hline
\end{tabular}\\
\end{minipage}
\end{center}
\end{table}

\subsubsection{Galactic distribution}

\begin{figure*}
\begin{center}
\includegraphics[width=0.45\textwidth, trim= 0 0 0 0]{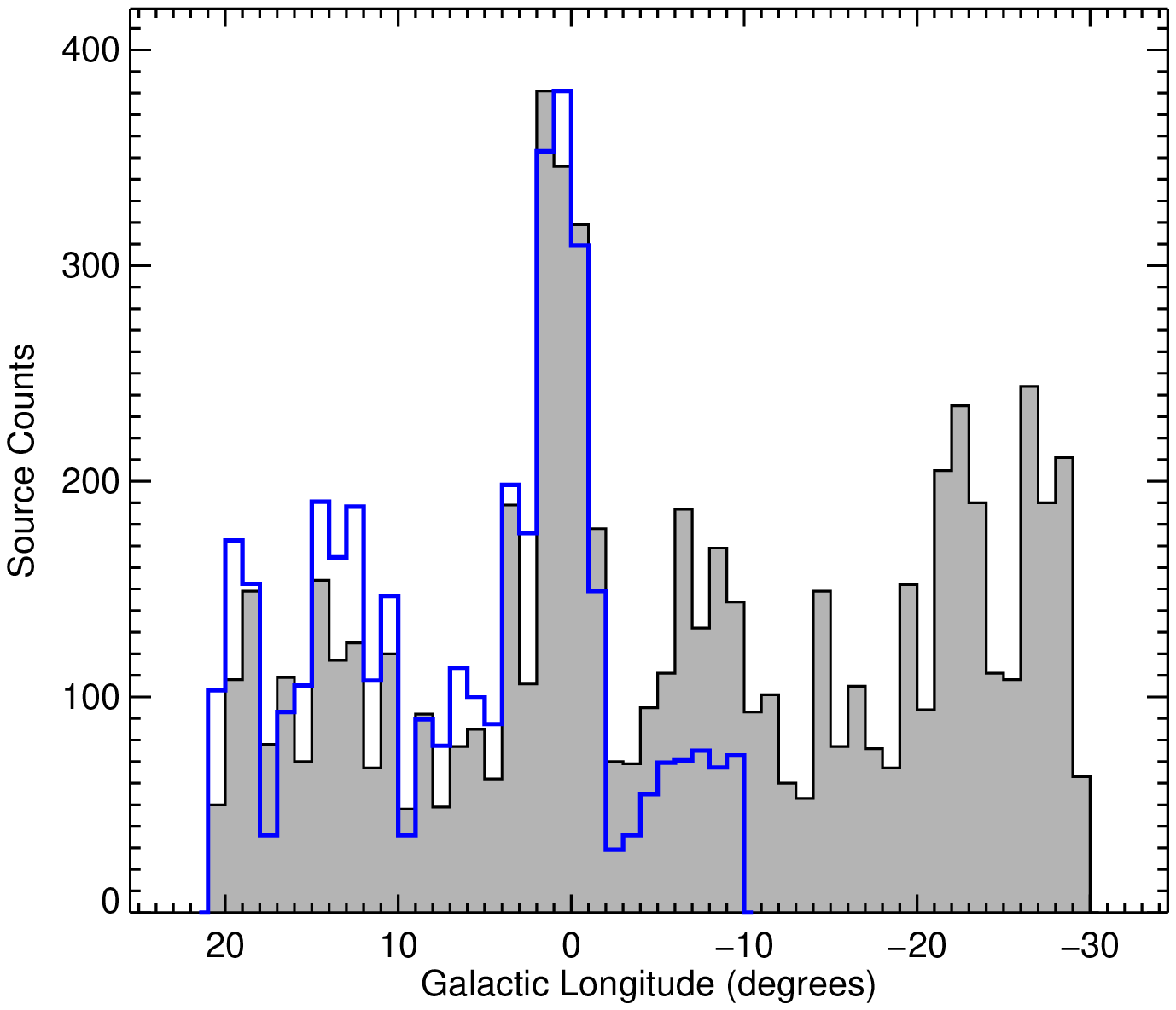}
\includegraphics[width=0.45\textwidth, trim= 0 0 0 0]{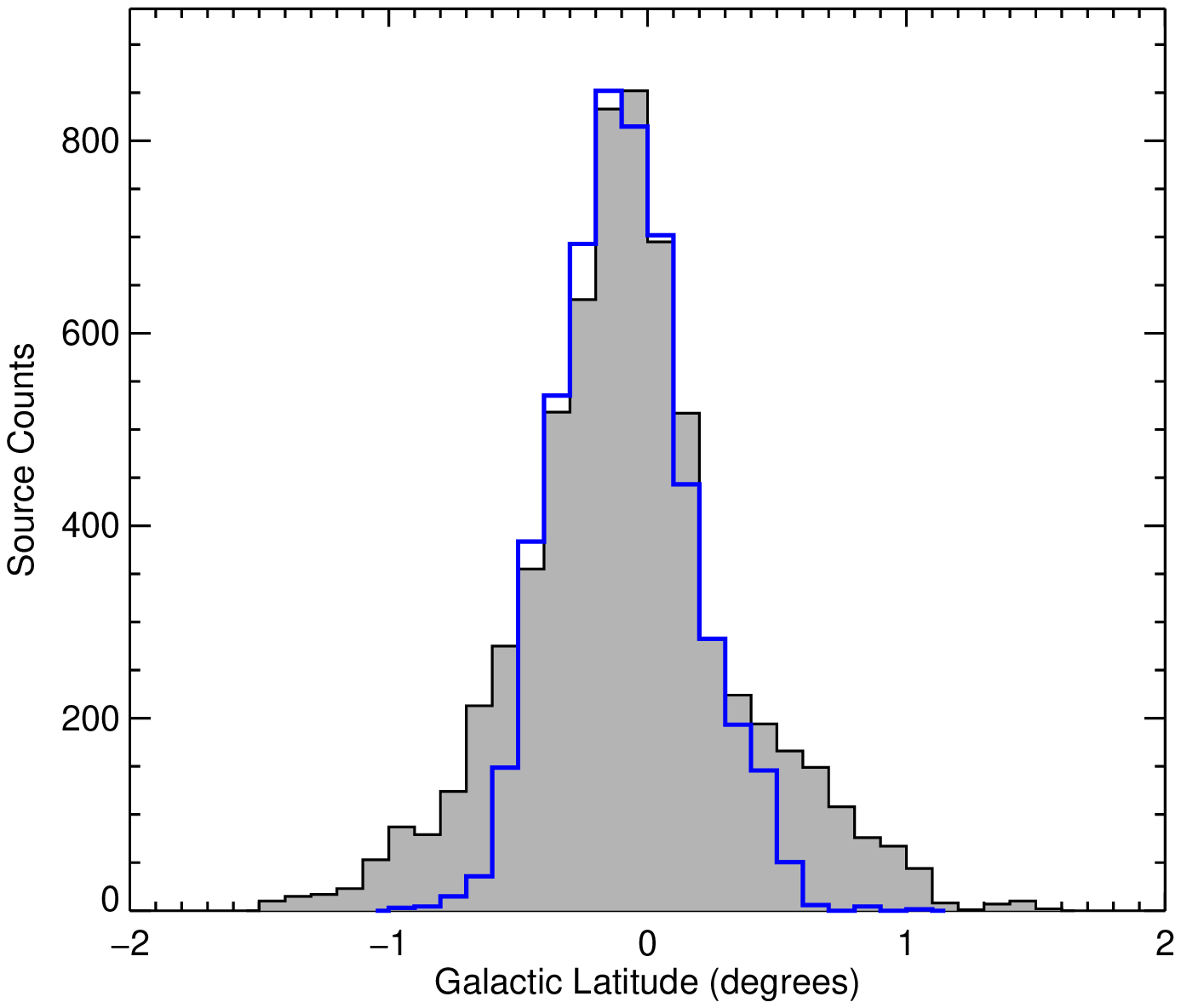}

\caption{\label{fig:gal_distribution} Histograms of the Galactic longitude and
  latitude distribution for ATLASGAL sources extracted by \sex\ (grey filled
  histogram) and the BGPS (blue histogram). In both plots the peak of the BGPS
  distribution has been normalised to the peak of the ATLASGAL
  distribution. The bin sizes used for the longitude (left panel) and latitude
  (right panel) distributions is 1\degr\ and 0.1\degr, respectively.}

\end{center}
\end{figure*}

\begin{figure*}
\begin{center}
\includegraphics[width=0.45\textwidth, trim= 0 0 0 0]{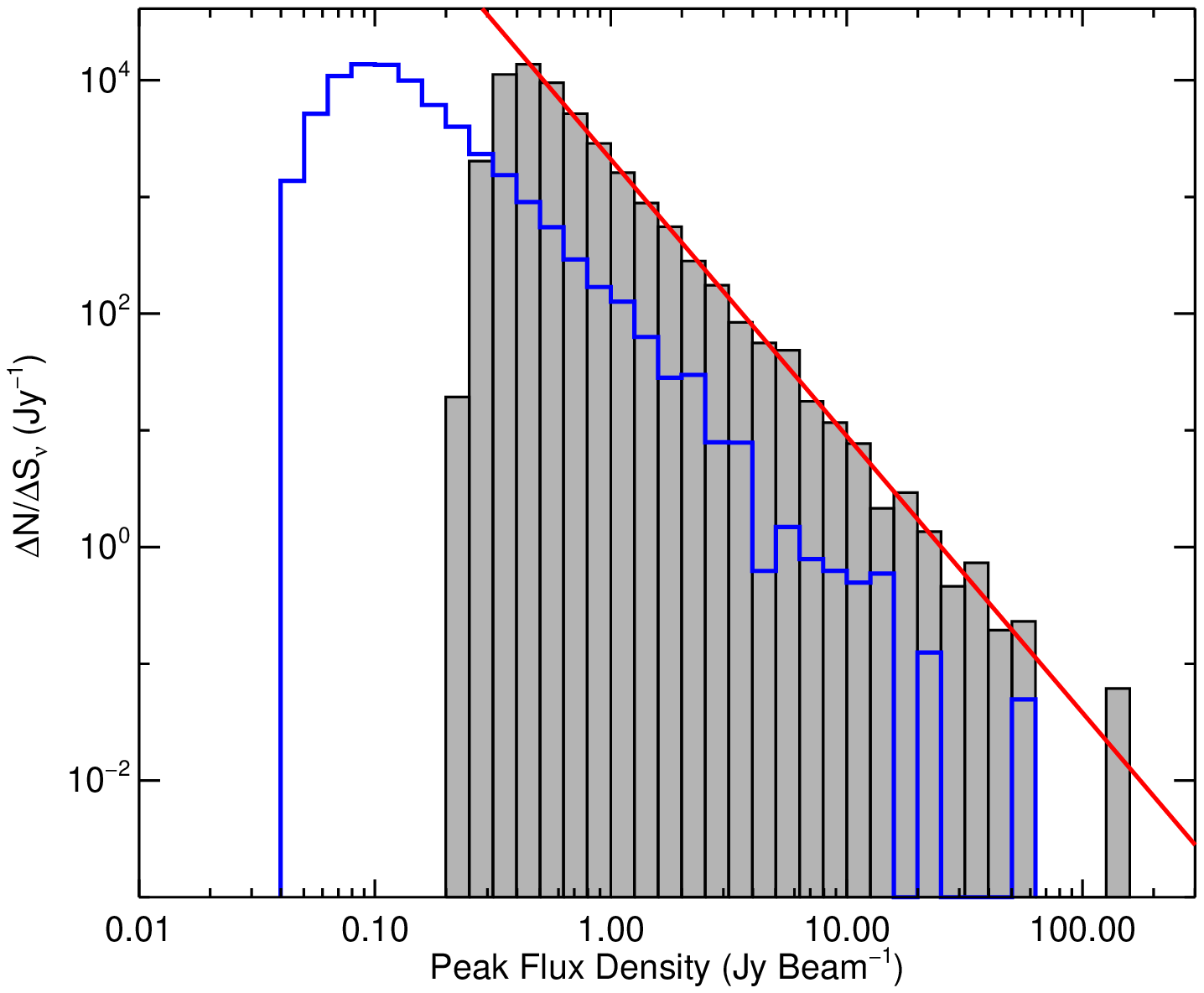}
\includegraphics[width=0.45\textwidth, trim= 0 0 0 0]{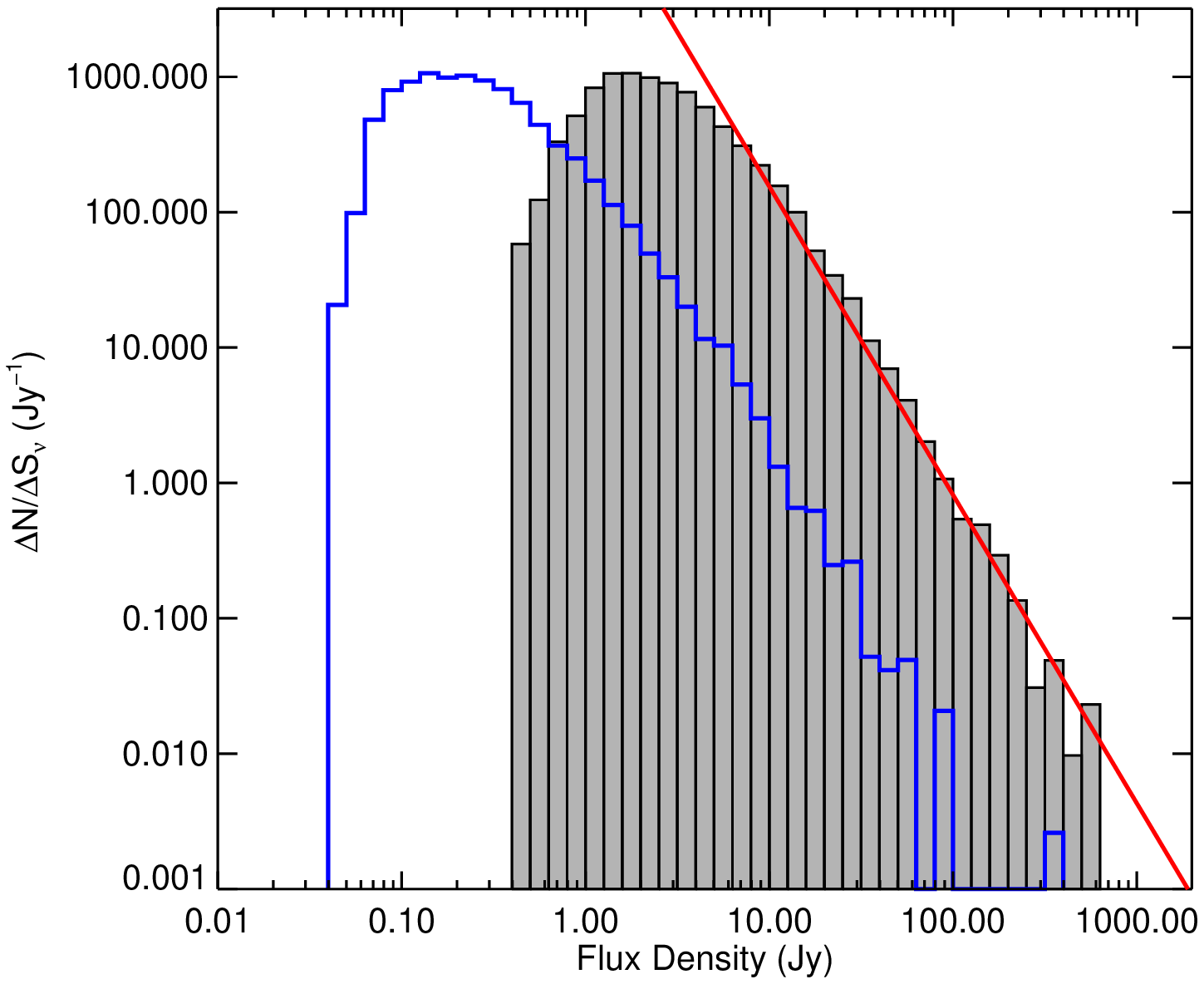}

\caption{\label{fig:flux_density} Flux density number distribution for
  ATLASGAL sources extracted by \sex\ (grey filled histogram) and the BGPS
  (blue histogram). In both plots the peak of the BGPS distribution has been
  normalised to the peak of the ATLASGAL distribution. In the left and right
  panels we present histograms of the peak and integrated flux densities
  measured for each source respectively. The red line on each plot shows the
  result of a linear least-squares fit to flux densities larger than the peak
  of each distribution.}

\end{center}
\end{figure*}

In Fig.\,\ref{fig:gal_distribution} we present histograms showing the
Galactic longitude and latitude distribution of the ATLASGAL catalogue
(left and right panels, respectively). Inspection of the longitude
distribution reveals a number of peaks in the source counts; these are
located at approximately $-$27\degr, $-$22\degr, $-$8\degr, 0\degr\
with a weaker peak at 14\degr. The Galactic structure traced by
ATLASGAL sources was the subject of a recent paper by
\citet{beuther2012}. The authors used the \texttt{CLUMPFIND} source
extraction algorithm to produce a source catalogue and they analysed
the distribution in Galactic longitude. They found that the positions
of the peaks in source counts coincide with the positions of the
tangent of spiral arms and individual star forming
  regions. Identifying the peaks in the source counts with the tangent
  of the Norma arm ($-$28\degr), the farside of the Near 3-kpc arm
($-$22\degr), the NGC6334/NGC6357 star forming regions ($-$8\degr),
the Galactic centre region and finally M16 and M17 regions (+15\degr),
respectively.

Comparing the longitude distribution of ATLASGAL sources with the BGPS
distribution we see that they are similar, both showing the same peaks,
however, we note that the ratio of the Galactic centre peak to the other peaks
is quite different. This is likely due to the different spatial resolution of
the two surveys. The complex emission seen towards the Galactic centre is more
easily deblended in the ATLASGAL data due to the higher spatial resolution
which results in a higher number of sources detected in this
region. Conversely, the lower resolution BGPS survey is more sensitive to
lower surface brightness structures resulting in a higher proportion of
sources being found away from the Galactic centre.

The latitude distribution shown in the right panel of
Fig.\,\ref{fig:gal_distribution} peaks significantly below 0$^\circ$
as also noted by \citet{Schuller} and more recently by
\citet{beuther2012}. This offset is consistent with the location
  of the Sun above the Galactic plane \citep{humphreys-1995}.
Comparing the latitude distribution to that of the BGPS we find them
again to be similar. Both ATLASGAL and BGPS distributions have the
same FWHM. We note that the tails of the distributions appear to be
significantly different, however, the BGPS latitude range is $|b| <
0.5\degr$ with only a few excursions to larger value of $|b|$ and so
the distributions are not really comparable for latitudes $|b| > 0.5$\degr.

\subsubsection{Flux distribution}

In Fig.\,\ref{fig:flux_density} we show the peak and integrated flux
distributions for the ATLASGAL sources (grey filled histogram) and the BGPS
sources located in the overlap region (blue histogram). These distributions
show the differential flux density spectrum (i.e., dN/d$S\nu$ $\simeq \Delta
\rm{N}/\Delta S\nu \propto S^{-\alpha}$; \citealt{rosolowsky2010}). Comparing
the distributions of the peak and integrated fluxes for the two catalogues
they seem to be very similar, however, with the higher frequency measurements
being shifted to higher values.

The flux distributions above the turnover can be approximated by a
power-law that extends over three orders of magnitude. Fitting these
parts of the distributions with a linear least-squares fit we find
that the peak and integrated distributions can be represented by the
values of $\alpha = 2.4\pm0.04$ and $2.3\pm0.06$, respectively. The
results of these fits are indicated by the red lines overplotted on
the left and right panels of Fig.\,\ref{fig:flux_density}. The
  values obtained from the same fit to the BGPS peak and integrated
  flux distributions are 2.4 and 1.9, respectively
  (\citealt{rosolowsky2010}). We note that the fit to the integrated
  fluxes is significantly steeper for the ATLASGAL catalogue than
  determined for the BGPS catalogue. This probably results from the
  different resolutions between with two surveys as the lower
  resolution BGPS survey is likely to blend smaller clumps, that are
  resolved in ATLASGAL, together into larger clumps with higher
  integrated fluxes. This leads to a higher proportion of the BGPS
  sources being found at higher fluxes and the shallower slope than found
  for the ATLASGAL catalogue.

The difference in the slope between the peak and integrated flux densities
results from the fact that point sources dominate the peak distribution
towards the completeness limit, whereas brighter sources tend to be larger and
thus have larger total flux densities than for the smaller sources. As pointed
out by \citet{rosolowsky2010}, this makes the integrated flux density
distribution top-heavy and effectively lowers the magnitude of the exponent.

Another interesting feature of the peak distribution is that the main body of
the distribution drops off completely at $\sim$70\,Jy\,beam$^{-1}$, however,
there is another outlying peak located at $\sim$150\,Jy\,beam$^{-1}$. The
bright sources that contribute to this flux density bin are the brightest
870\,$\mu$m sources in the catalogue and form part of a dust ridge associated
with the Sgr B2 high-mass star forming region.

\subsubsection{Angular size distribution}

\begin{figure}
\begin{center}

\includegraphics[width=0.45\textwidth, trim= 0 0 0 0]{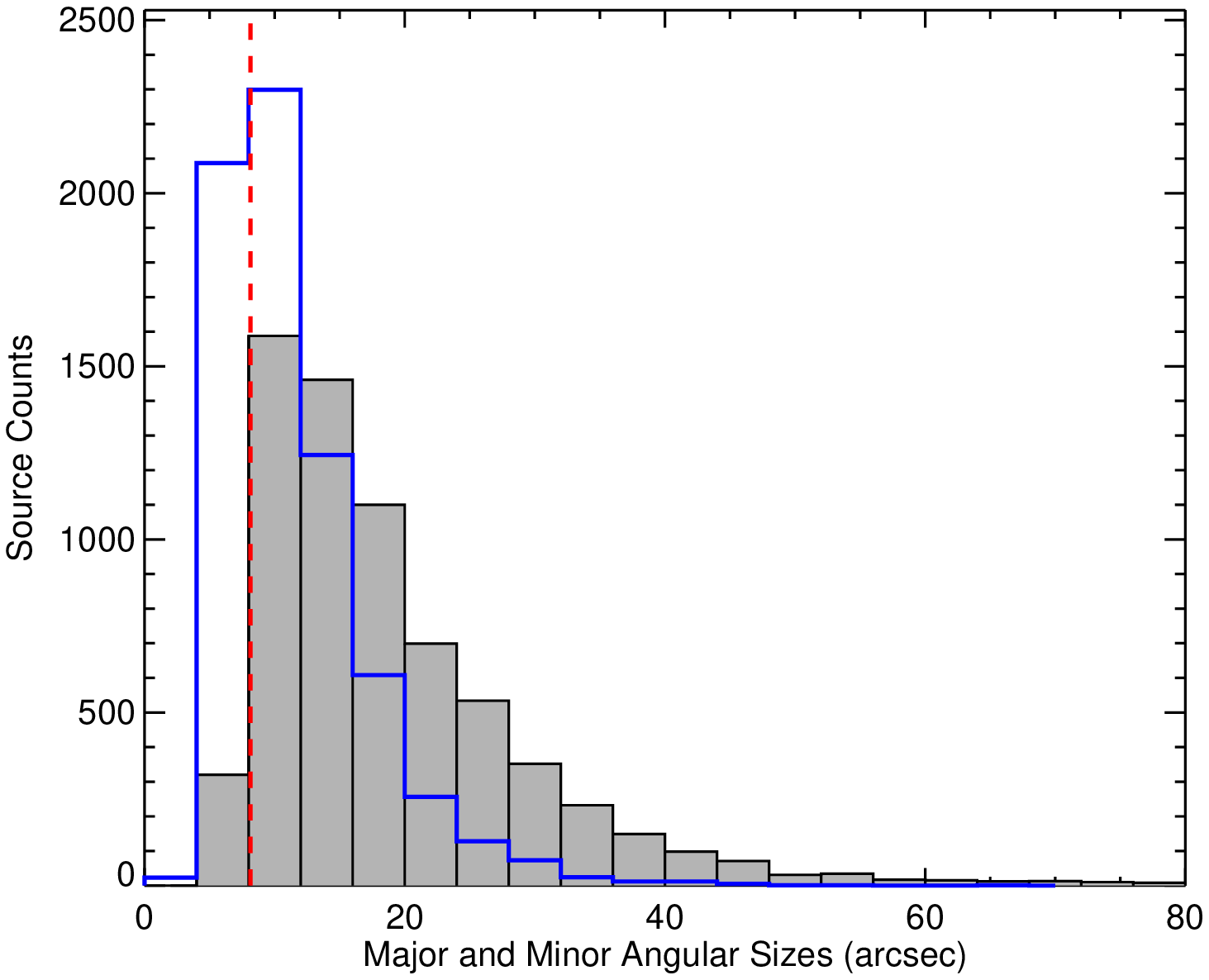}
\includegraphics[width=0.45\textwidth, trim= 0 0 0 0]{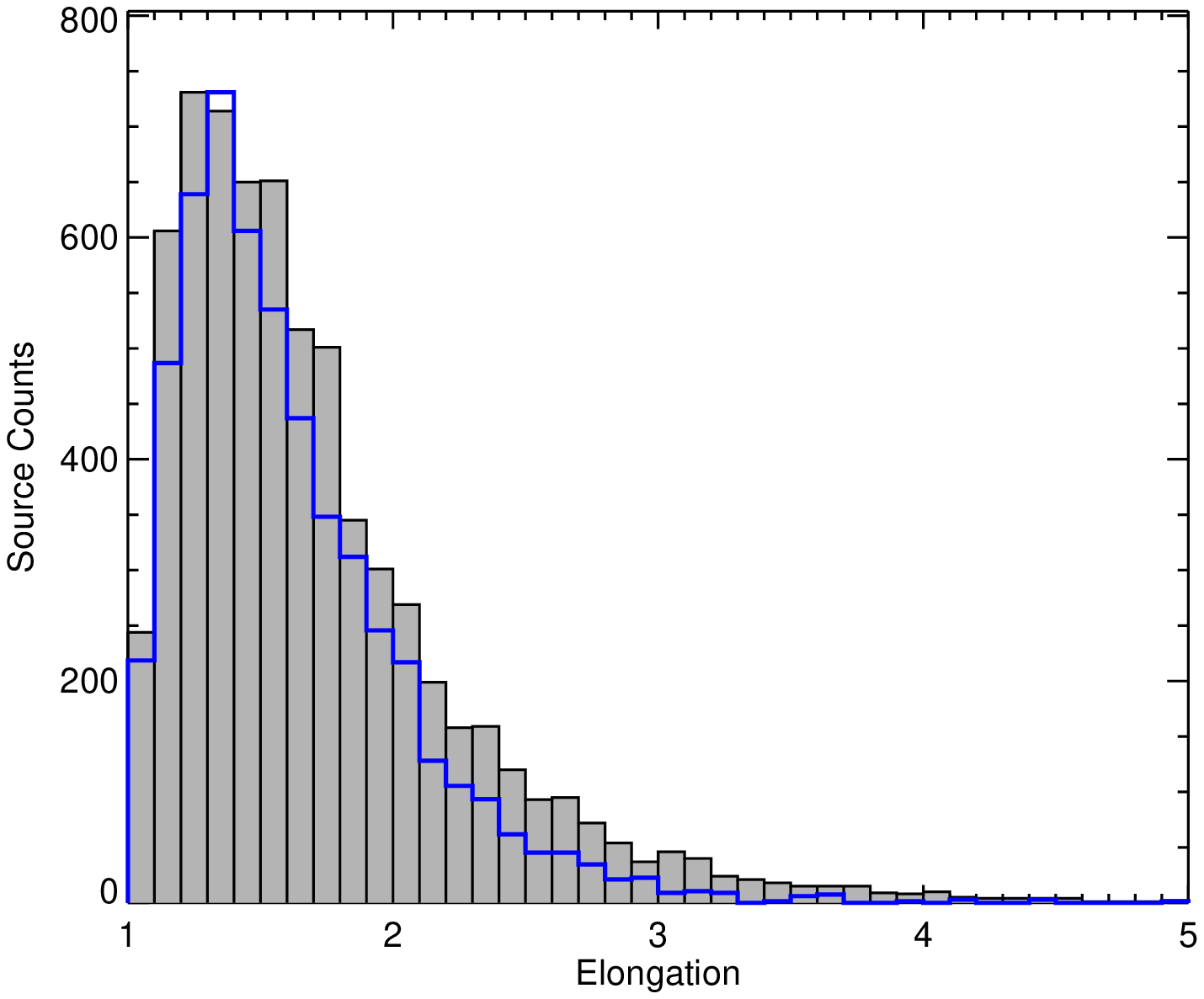}
\includegraphics[width=0.45\textwidth, trim= 0 0 0 0]{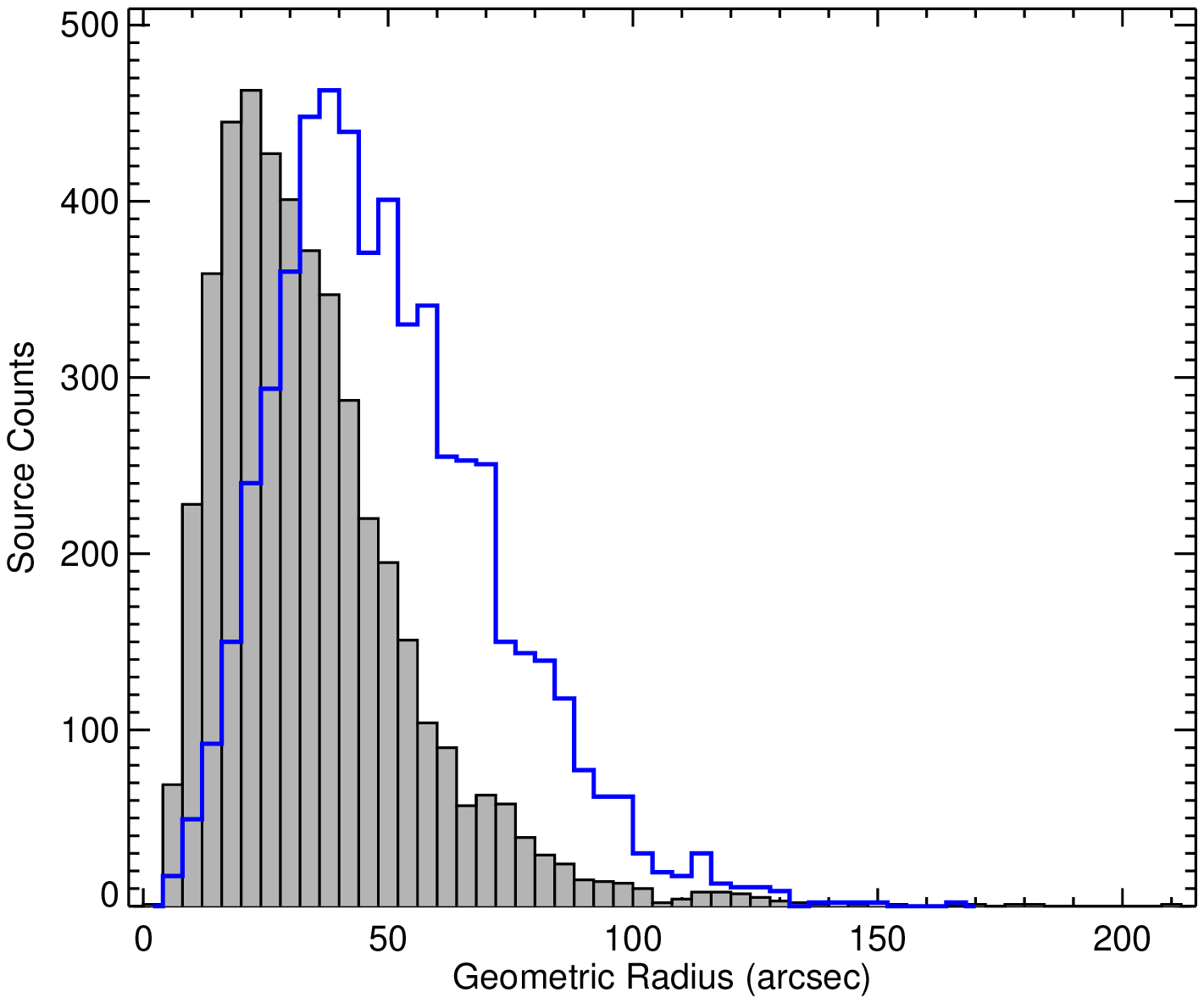}

\caption{\label{fig:major_minor_distribution} Histograms of the
  angular size distribution of detected ATLASGAL sources. In the upper
  panel we show the distribution of the semi-major (filled grey
  histogram) and semi-minor (blue open histogram) size distributions
  of detected ATLASGAL sources. The red dashed line shows the ATLASGAL 
  \rms\ beam size . The bin size used for both parameters is 4\arcsec. In the
  middle and lower panels we present histograms of the elongation and
  effective radii of ATLASGAL sources (grey filled histogram) and the
  BGPS (blue histogram). In both cases the peak of the BGPS
  distribution has been normalised to the peak of the ATLASGAL
  distribution. The bin sizes used for the elongation and radii
  distributions are 0.1 and 4\arcsec\ respectively.}

\end{center}
\end{figure}

In Fig.\,\ref{fig:major_minor_distribution} we present plots showing
the size distributions of the catalogued sources. In the upper panel
of this figure we show histograms of the semi-major
($\sigma_{\rm{maj}}$ ) and semi-minor ($\sigma_{\rm{min}}$)
distributions and the ATLASGAL beam size ($\sigma_{\rm{bm}} \simeq
8''$, red dotted line). We note that there is a significant number of
sources with sizes smaller than the beam, particularly with regard to
the semi-minor axis. This is due to the fact that the sizes are
derived from moment measurements of the emission distribution above
the threshold value ($\sim$3$\sigma_{\rm{\rms}}$) and therefore do not
take account of all of the emission associated with each source down
to the zero-intensity level. As previously noted by
\citet{rosolowsky2010} this has a disproportionate effect on weaker
sources and often results in their sizes being underestimated. We find the mean and median values of the peak fluxes for sources in our catalogue that are smaller than the beam are 0.52 and 0.43\,Jy\,beam$^{-1}$, respectively, and therefore this sample is dominated by sources around the sensitivity limit with only a few percent having peak fluxes above 1\,Jy\,beam$^{-1}$.

In the middle panel of Fig.\,\ref{fig:major_minor_distribution} we present a
histogram showing the elongation distribution, that is the ratio of semi-major
to semi-minor axes. The distribution has a peak and median value of 1.3 and
1.6, respectively, which indicates that most sources in the catalogue are
significantly elongated. The elongation distribution closely matches that
found for the BGPS sources. The only noticeable difference is that the
ATLASGAL catalogue tends to have a slightly higher degree of elongation,
possibly a result of the higher resolution of ATLASGAL survey. However, both
catalogues appear to contain a significant number of very elongated sources
reflecting the fundamentally filamentary structure of molecular clouds.

In the lower panel of Fig\,\ref{fig:major_minor_distribution} we present the
effective radii of the ATLASGAL and BGPS detections. The two distributions
again appear to be broadly similar, however, the whole BGPS source
distribution appears to have been shifted 10-15\arcsec\ to the right of the
ATLASGAL distribution, which is clearly a result of the different survey
resolutions.

\section{Comparison with other Galactic plane surveys}
\label{cross}

\subsection{Infrared catalogues}

We performed a cross correlation of our catalogue with the
far-infrared {\em IRAS} Point Source Catalogue
(\citealt{iras_catalog}), and with the mid-infrared Midcourse Space
eXperiment ({\em MSX}) Point Source Catalogue (\citealt{msx_catalog}).
We used a search radius of 30\arcsec\ around the peak flux position
for all these survey catalogues. The resolution of the {\em MSX}
survey and the APEX telescope used for ATLASGAL are very similar
(18\arcsec\ and 19.2\arcsec, respectively) making the correlation
between both data sets relatively straightforward. This
  represents an initial attempt to associate ATLASGAL sources with their
  IR counterpart, and further associations with other catalogues
(e.g. those extracted from the {\em Spitzer} GLIMPSE and MIPSGAL
surveys) will be addressed in a future paper.

Infrared emission detected by the {\em IRAS} and {\em MSX} surveys generally
implies that star formation is already underway within a clump, therefore,
these surveys trace objects in a relatively evolved stage of star formation.
The ATLASGAL catalogue presented here contains \nsources\ sources, of which we
found \niras\ associated with {\em IRAS} sources and \nmsx\ associated with
{\em MSX} point sources. The number of sources with {\em IRAS} and {\em MSX}
association is \nall.

We found that \nnothing\ objects, corresponding to 62\% of the
ATLASGAL sources, are not associated with any mid- or far-infrared
point sources found in the {\em MSX} or {\em IRAS} catalogues. These
objects are excellent candidates for cold dense clumps in an early
stage of evolution, prior to the birth of high-mass
protostars. Further correlation of these sources with other infrared
surveys such as GLIMPSE or MIPSGAL, can give further information on
their evolutionary stage, based on the presence or absence of star
formation indicators such as green fuzzies (enhancement at
4.5\,$\mu$m, often referred to as extended green objects, e.g.,
\citealt{cyganowski2008}) or 24\,$\mu$m point sources. Moreover,
follow up studies of these cold dense clumps in molecular lines and
their detailed study with next generation telescopes such as the ALMA
and the Jansky VLA will help improve our understanding the initial
stages of high-mass star formation.

\begin{figure}
  \centering \includegraphics[width=0.45\textwidth, trim= 0 0 0 0]{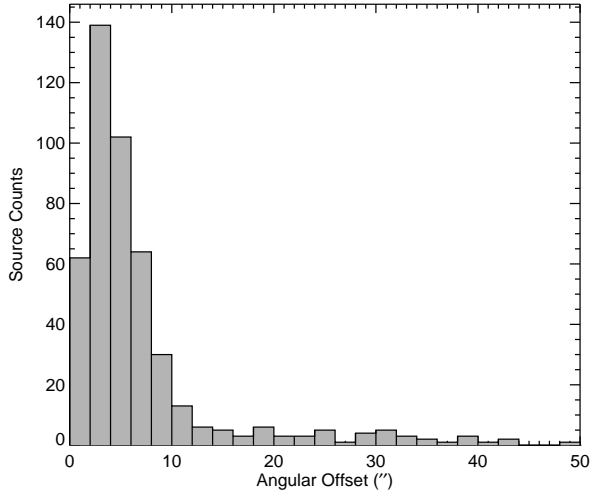}
\caption{\label{fig:mmb_atlas_offset} Distribution of separation between methanol maser
  sources from the MMB survey and ATLASGAL sources. This plot has been truncated at an angular offset of 50\arcsec, however, only 7 ATLASGAL-MMB associations have larger offsets than this. The bin size is 2\arcsec.}
\end{figure}

\subsection{Methanol masers}

Methanol masers are well-known probes of early phases of high-mass star
formation, in particular the strong 6.7\,GHz Class-II maser \citep{menten1991}.
Recently a Galactic plane survey of this maser transition using a 7-beam
receiver on the Parkes telescope has been completed \citep{green+2009}. These
maser detections were followed up at high angular-resolution with the
Australian Telescope Compact Array to obtain sub-arcsecond positional
accuracy. To date the MMB has reported the positions of 707 methanol masers
located between $186\degr < \ell\ < 20\degr$ and $|b| < 2\degr$
(\citealt{caswell2010,green2010,caswell2011,green2012}), of these 493 are
located within ATLASGAL fields presented in this paper (i.e., 330\degr $ <
\ell <$ 21\degr). Cross-matching these MMB sources with the ATLASGAL catalogue
generated from the \sex\ algorithm we find 471 positional associations, which
corresponds to $\sim$96\% of the available MMB sources but only $\sim$7\% of
the ATLASGAL catalogue.

In Fig.\,\ref{fig:mmb_atlas_offset} we show the angular separation
between the ATLASGAL-MMB associations. The sharp peak seen towards
small angular offsets reveals a strong correlation between the
position of the methanol maser and the peak column density of each
clump as traced by the submm continuum emission. We find that
$\sim$87\% of all ATLASGAL-MMB associations are found with angular
separations $<$\,12\arcsec, after which the distribution of the
offsets flattens to an almost constant background level. We find that 61 ATLASGAL-MMB associations have angular offsets larger than 12\arcsec; this could
indicate the presence of clumpy substructure that is smaller than the
size of the APEX beam or a chance alignment with an MMB source
associated with an undetected dust clump.

Inspection of the ATLASGAL maps at the locations of the 22\,MMB sources not
found to be associated with an ATLASGAL source reveals that nearly all of
these masers are associated with either weak diffuse emission, which would
have been filtered out by the background subtraction, or weak compact emission
with sizes smaller than that of the APEX beam. Since methanol masers are
almost exclusively associated with high-mass star forming regions it is very
likely that the majority of these masers are located at the far side of the
Galaxy. These masers have peak fluxes between 0.64 and 3.99\,Jy\,beam$^{-1}$,
well above the MMB surveys sensitivity of 0.17\,Jy\,beam$^{-1}$, which would
suggest that the methanol maser surveys can reveal some high-mass star forming
regions that are missed at the sensitivity of ATLASGAL. 

A detailed analysis of the correlation and anti-correlation between the ATLASGAL and MMB catalogues and their derived properties is beyond the scope of this paper, but will be presented in a subsequent paper \citep{Urquhart-2012}.

\subsection{Ultra-compact {\sc Hii} regions}

We used the unbiased 5\,GHz VLA radio continuum survey by \citet{ref-bwhz} to study the association between
UC{\sc Hii} regions and submm clumps. In this VLA survey, the Galactic range from
$\ell=350-40\degr$ and $|b| < 0.4\degr$ was covered with the C and CnD
configurations, resulting in an angular resolution of $\sim$4\arcsec.  This radio catalogue contains $\sim$1300 sources, 30\% of which they tentatively identify as UC{\sc Hii} regions. 

Using a search radius of 20\arcsec\ we find in the overlap region of both surveys 170 compact ATLASGAL sources that are positionally coincident with a radio continuum emission ($\sim$6\% of the ATLASGAL sample). Turning to the radio sources we find that 23\% of these are associated ATLASGAL sources and are therefore likely to be genuine UC{\sc Hii} regions, which is in reasonable agreement with \citet{ref-bwhz} initial estimate.

\subsection{Infrared Dark Clouds}

A considerable fraction of submillimetre bright sources appear dark in Spitzer
infrared images. Using the \citet{ref-peretto} catalogue of infrared dark
clouds (IRDC) that is based on the Spitzer GLIMPSE images, we determined the
association of ATLASGAL clumps with IRDCs. An association was based on overlap
of the source ellipses in the two catalogues. In the part of the Galaxy common
to both catalogues, we find that 30\% of the IRDCs are associated with
ATLASGAL clumps. A similar analysis was recently conducted by \citet{wilcock2012} in a complementary part of the Galaxy ($\ell=300$-330\degr) who
determined how many of the IRDCs were seen in the Herschel Hi-GAL
250/500\,$\mu$m bands. They found a fraction of 38\% which is, given the
higher mass sensitivity of the Herschel survey, consistent with the ATLASGAL
result.

\begin{figure}[ht]
  \centering
  
  \includegraphics[width=0.49\textwidth]{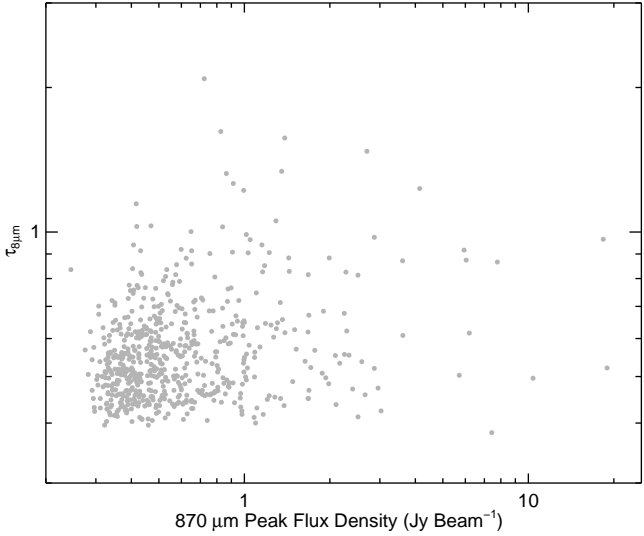}
  \caption{\label{fig-irdc} Comparison between peak submm flux
    densities and 8\,$\mu$m peak optical depths for ATLASGAL sources
    associated with IRDCs from the \citet{ref-peretto} catalogue.}
\end{figure}

The IRDC peak optical depths are plotted against the peak flux densities
measured in ATLASGAL in Fig.\,\ref{fig-irdc}. Since both quantities are
proportional to column density, a correlation is expected. The main
uncertainty in the determination of column densities from ATLASGAL data is the
dependence of the flux on the dust temperature, which might vary by about a
factor 2 (based on ammonia temperatures of ATLASGAL sources, \citealt{wienen2012}). Large uncertainties in the determination of IR optical depths result
from the estimation of the foreground emission and the subtraction of the
background. In addition, for high optical depths, the extinction in the IR
will saturate \citep[e.g.][]{Vasyunina2009}. Given these major uncertainties of the latter method, column
density estimates by optically thin dust emission in the (sub)millimetre still
provide the most reliable method to measure the high column densities towards
high-mass star forming clumps.

\subsection{MSX 21~$\mu$m  association}

To study the mid-infrared properties of the ATLASGAL compact sources in more
detail, we extracted the 21\,$\mu$m flux densities from the MSX survey \citep{price2001} towards the ATLASGAL peaks. For such an analysis, the MSX data
offer two advantages compared to the more sensitive and higher angular
resolution data from the Spitzer MIPSGAL 24\,$\mu$m survey (Carey et al. 2009):
(i) the spatial resolution of the MSX data (18.3\arcsec) matches the ATLASGAL
resolution extremely well and (ii) the MSX data do not suffer from saturation
that MIPSGAL shows towards strong sources, so that the full range of MIR
brightnesses can be included in the analysis. This comparison is also more
complete than a comparison with the MSX point source catalogue alone, since
many ATLASGAL sources are  associated with more complex MIR emission not
covered in the MSX point source catalogue. 

\begin{figure}
  \centering
  \includegraphics[width=0.49\textwidth]{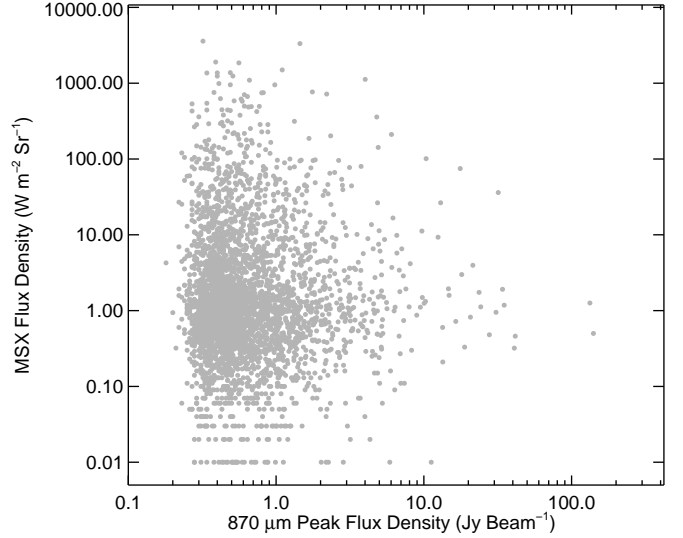}
\caption{\label{compare_msx} Comparison of peak 870\,$\mu$m fluxes of ATLASGAL sources and the corresponding MSX 21\,$\mu$m fluxes at the same position.}
\end{figure}

In Fig.\,\ref{compare_msx} we show the correlation of 870\,$\mu$m peak
flux density of the ATLASGAL sources and the 21\,$\mu$m MSX fluxes
measured at the same position. To remove the contribution from
  the variable background seen in the 21\,$\mu$m images we have
  averaged the emission in an annulus centred on the ATLASGAL position
  with inner and outer radius of 30\arcsec\ and 120\arcsec,
  respectively, and subtracted this from the 21\,$\mu$m flux measured
  towards the centre of the ATLASGAL sources. The MSX fluxes span a
  large range and only a loose correlation of increasing MSX with
  ATLASGAL flux is seen. In fact, a large fraction of the ATLASGAL
  sources ($\sim$50\%) show no 21\,$\mu$m emission above the local
  background level, confirming again that many of the ATLASGAL compact
  sources do not have embedded or associated luminous infrared
  sources. However, the fact that approximately half of the ATLASGAL sources
  have MSX 21\,$\mu$m emission would suggest that the lower limit for
  the number associated with star formation is closer to 50\%. This
  fraction is significantly higher than found from the straight
  comparison with the IRAS and MSX PSCs preformed in Sect.\,6.1. This
  fraction is likely to increase when the ATLASGAL catalogue is
  compared with the Spitzer GLIMPSE and MIPSGAL surveys, which will be
  presented in a future paper.

\begin{figure}
  \centering
\includegraphics[width=0.49\textwidth]{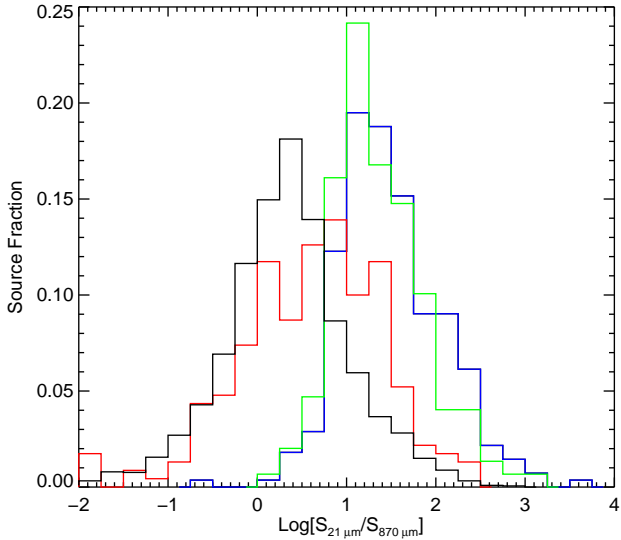}
    \caption{\label{compare_msx_samples} Histogram plots showing the
    distribution of MSX 21\,$\mu$m over ATLASGAL 870\,$\mu$m flux ratio for
    whole ATLASGAL catalogue (black), the methanol masers identified by the
    MMB survey (red) and UC{\sc Hii} regions and YSOs identified by the RMS survey
    (coloured blue and green, respectively).}
\end{figure}

These background corrected MSX 21\,$\mu$m fluxes can be used to compare
the mid-infrared properties of different subsamples ATLASGAL sources.
In general terms, the ATLASGAL flux is a measure of the total mass of
the clumps while the 21\,$\mu$m MSX flux probes a warm component
($\sim$100\,K) of their spectral energy distributions. Therefore the
ratio of both fluxes is a crude measure of the fraction of heated
material. We show the histogram of this ratio in
Fig.~\ref{compare_msx_samples} and compare the distribution of this
ratio with the corresponding histograms for sources from the RMS
survey \citep{urquhart2007c}, both for identified high-mass young
stellar objects and HII regions, and with the histogram of this ratio
for CH$_3$OH masers from the MMB survey \citep{caswell2010}. The peak
of the distribution for the RMS YSOs and HII regions is very
comparable (green and blue histograms, respectively), and it is
clearly offset from the ATLASGAL peak, again showing that with
ATLASGAL many clumps with either no star formation or the very early stages of star formation are probed. The 21\,$\mu$m/870\,$\mu$m flux ratio for the methanol
masers has a broader distribution (red histogram) which is consistent
with covering the flux ratios from both, the bulk of the ATLASGAL and
the RMS sources.

\begin{table}
  \begin{center}\caption{\label{tbl:assoc} Summary of associations between ATLASGAL compact sources and infrared and submillimetre surveys.}
    \begin{minipage}{\linewidth}
      % \scriptsize
      \begin{tabular}{lccr}
        \hline \hline
        \multicolumn{1}{l}{Catalogue}  &  \multicolumn{1}{c}{Coverage within ATLASGAL}&  \multicolumn{1}{c}{N sources}\\
        \hline
        IRAS    & Full survey & 110\\
        MSX     & Full survey & 1669\\
        IRAS-MSX&Full survey & 645\\
         MMB     & $-15^\circ<l<20^\circ$;$|b|<2^\circ$ &  459 \\
        UC{\sc Hii} regions  & $10^\circ<l<21^\circ$;$|b|<1^\circ$&  170 \\
        MSX 21$\mu$m$^*$&Full survey & 3410 \\
        \hline
      \end{tabular}\\
    \end{minipage}
   
  \end{center}
   $^*$ Number of ATLASGAL sources associated with MSX 21\,$\mu$m emission above the local background emission (see Sect.\,6.5 for details).
\end{table}

\section{Summary}
\label{sum}

The ATLASGAL has mapped 420\,deg$^2$ of the inner Galaxy at
870\,$\mu$m, which traces thermal emission from dense dust clumps. Here
we present a compact source catalogue covering 153\,deg$^2$ between
Galactic longitude $330\degr < \ell < 21\degr$ and $|b|<
$1.5\degr. The source extraction algorithm \sex\ has been used to
identify clumps and determine their parameters. To mitigate the
varying \rms\ noise across the fields we performed the source
detection on background subtracted signal-to-noise map, however, the
source parameters were determined from the original emission maps. In
total \nsources\ sources have been identified above a 3$\sigma$
threshold ($\sim$200\,mJy\,beam$^{-1}$) and we find the catalogue is
99\% complete above 6$\sigma$, which corresponds to
$\sim$0.4\,Jy\,beam$^{-1}$.

Comparing this catalogue of dense clumps to the $MSX$ and $IRAS$ point
source catalogues we find $\sim$40\% are associated with far- and
mid-infrared compact sources. The infrared emission likely arises from
embedded young stellar objects (YSOs) or UC{\sc Hii} regions and is a
strong indication that star formation is currently underway and is in
an evolved phase in these clumps. This is supported by the positional
correlation of 5\,GHz radio continuum emission with approximately 6\%
of these mid-infrared associated sources. In addition to the
mid-infrared and radio associations we find a further $\sim$500 clumps
are associated with methanol masers ($\sim$7\%). Methanol masers are
thought to trace an earlier evolutionary stage than the mid-infrared
emission and have been almost exclusively found to be associated with
high-mass star formation regions. A summary with the number of
  sources with associations among the different surveys is shown in
  Table\,\ref{tbl:assoc}.

Combined the infrared, radio continuum and methanol maser associated
sources make up approximately 50\% of the catalogue. These include all
of the important evolutionary stages from young deeply embedded
protostars identified by methanol masers to mid-infrared bright YSOs
and UC{\sc Hii} regions and thus allows for the first time these different
samples to be placed into a global framework. However, possibly more
intriguing is the other 50\% of the sample towards which none of the
star formation tracers have been detected as these sources are likely
to be in an early stage of evolution. These sources are excellent
targets for follow up observation, e.g. in the near future at very
high spatial resolution with
ALMA.

\begin{acknowledgements}
   We thank the referee for their useful comments and suggestions 
    that have helped to clarify some important points in this paper. This
  work was partially funded by the ERC Advanced Investigator Grant
  GLOSTAR (247078) and was partially carried out within the
  Collaborative Research Council 956, sub-project A6, funded by the
  Deutsche Forschungsgemeinschaft (DFG). L.B. acknowledges support
  from CONICYT through project BASAL PFB-06. J. Tackenberg and
  M. Wienen are supported for this research through a stipend from the
  International Max Planck Research School (IMPRS).
\end{acknowledgements}

\bibliography{bibliografia}

\begin{thebibliography}{47}
\expandafter\ifx\csname natexlab\endcsname\relax\def\natexlab#1{#1}\fi

\bibitem[{{Aguirre} {et~al.}(2011){Aguirre}, {Ginsburg}, {Dunham}, {Drosback},
  {Bally}, {Battersby}, {Bradley}, {Cyganowski}, {Dowell}, {Evans}, {Glenn},
  {Harvey}, {Rosolowsky}, {Stringfellow}, {Walawender}, \&
  {Williams}}]{aguirre2011}
{Aguirre}, J.~E., {Ginsburg}, A.~G., {Dunham}, M.~K., {et~al.} 2011, \apjs,
  192, 4

\bibitem[{{Becker} {et~al.}(1994){Becker}, {White}, {Helfand}, \&
  {Zoonematkermani}}]{ref-bwhz}
{Becker}, R.~H., {White}, R.~L., {Helfand}, D.~J., \& {Zoonematkermani}, S.
  1994, \apjs, 91, 347

\bibitem[{{Benjamin} {et~al.}(2003){Benjamin}, {Churchwell}, {Babler}, {Bania},
  {Clemens}, {Cohen}, {Dickey}, {Indebetouw}, {Jackson}, {Kobulnicky},
  {Lazarian}, {Marston}, {Mathis}, {Meade}, {Seager}, {Stolovy}, {Watson},
  {Whitney}, {Wolff}, \& {Wolfire}}]{benjamin-2003}
{Benjamin}, R.~A., {Churchwell}, E., {Babler}, B.~L., {et~al.} 2003, \pasp,
  115, 953

\bibitem[{{Bertin} \& {Arnouts}(1996)}]{bertin}
{Bertin}, E. \& {Arnouts}, S. 1996, \aaps, 117, 393

\bibitem[{{Beuther} {et~al.}(2012){Beuther}, {Tackenberg}, {Linz}, {Henning},
  {Schuller}, {Wyrowski}, {Schilke}, {Menten}, {Robitaille}, {Walmsley},
  {Bronfman}, {Motte}, {Nguyen-Luong}, \& {Bontemps}}]{beuther2012}
{Beuther}, H., {Tackenberg}, J., {Linz}, H., {et~al.} 2012, \apj, 747, 43

\bibitem[{{Bonnell} {et~al.}(2004){Bonnell}, {Vine}, \& {Bate}}]{bonnell-2004}
{Bonnell}, I.~A., {Vine}, S.~G., \& {Bate}, M.~R. 2004, \mnras, 349, 735

\bibitem[{{Carey} {et~al.}(2009){Carey}, {Noriega-Crespo}, {Mizuno}, {Shenoy},
  {Paladini}, {Kraemer}, {Price}, {Flagey}, {Ryan}, {Ingalls}, {Kuchar},
  {Pinheiro Gon{\c c}alves}, {Indebetouw}, {Billot}, {Marleau}, {Padgett},
  {Rebull}, {Bressert}, {Ali}, {Molinari}, {Martin}, {Berriman}, {Boulanger},
  {Latter}, {Miville-Deschenes}, {Shipman}, \& {Testi}}]{carey-2009}
{Carey}, S.~J., {Noriega-Crespo}, A., {Mizuno}, D.~R., {et~al.} 2009, \pasp,
  121, 76

\bibitem[{{Caswell} {et~al.}(2010){Caswell}, {Fuller}, {Green}, {Avison},
  {Breen}, {Brooks}, {Burton}, {Chrysostomou}, {Cox}, {Diamond}, {Ellingsen},
  {Gray}, {Hoare}, {Masheder}, {McClure-Griffiths}, {Pestalozzi}, {Phillips},
  {Quinn}, {Thompson}, {Voronkov}, {Walsh}, {Ward-Thompson}, {Wong-McSweeney},
  {Yates}, \& {Cohen}}]{caswell2010}
{Caswell}, J.~L., {Fuller}, G.~A., {Green}, J.~A., {et~al.} 2010, \mnras, 404,
  1029

\bibitem[{{Caswell} {et~al.}(2011){Caswell}, {Fuller}, {Green}, {Avison},
  {Breen}, {Ellingsen}, {Gray}, {Pestalozzi}, {Quinn}, {Thompson}, \&
  {Voronkov}}]{caswell2011}
{Caswell}, J.~L., {Fuller}, G.~A., {Green}, J.~A., {et~al.} 2011, \mnras, 417,
  1964

\bibitem[{{Cyganowski} {et~al.}(2008){Cyganowski}, {Whitney}, {Holden},
  {Braden}, {Brogan}, {Churchwell}, {Indebetouw}, {Watson}, {Babler},
  {Benjamin}, {Gomez}, {Meade}, {Povich}, {Robitaille}, \&
  {Watson}}]{cyganowski2008}
{Cyganowski}, C.~J., {Whitney}, B.~A., {Holden}, E., {et~al.} 2008, \aj, 136,
  2391

\bibitem[{{Deharveng} {et~al.}(2010){Deharveng}, {Schuller}, {Anderson},
  {Zavagno}, {Wyrowski}, {Menten}, {Bronfman}, {Testi}, {Walmsley}, \&
  {Wienen}}]{deharveng+2010}
{Deharveng}, L., {Schuller}, F., {Anderson}, L.~D., {et~al.} 2010, \aap, 523,
  A6

\bibitem[{{Di Francesco}(2008)}]{di_francesco2008}
{Di Francesco}, J. 2008, in Bulletin of the American Astronomical Society,
  Vol.~40, American Astronomical Society Meeting Abstracts \#212, 271

\bibitem[{{Dunham} {et~al.}(2011){Dunham}, {Rosolowsky}, {Evans}, {Cyganowski},
  \& {Urquhart}}]{dunham2011}
{Dunham}, M.~K., {Rosolowsky}, E., {Evans}, II, N.~J., {Cyganowski}, C., \&
  {Urquhart}, J.~S. 2011, \apj, 741, 110

\bibitem[{{Egan} {et~al.}(2003){Egan}, {Price}, {Kraemer}, {Mizuno}, {Carey},
  {Wright}, {Engelke}, {Cohen}, \& {Gugliotti}}]{msx_catalog}
{Egan}, M.~P., {Price}, S.~D., {Kraemer}, K.~E., {et~al.} 2003, VizieR Online
  Data Catalog, 5114, 0

\bibitem[{{Foster} {et~al.}(2011){Foster}, {Jackson}, {Barnes}, {Barris},
  {Brooks}, {Cunningham}, {Finn}, {Fuller}, {Longmore}, {Mascoop}, {Peretto},
  {Rathborne}, {Sanhueza}, {Schuller}, \& {Wyrowski}}]{Foster-2011}
{Foster}, J.~B., {Jackson}, J.~M., {Barnes}, P.~J., {et~al.} 2011, \apjs, 197,
  25

\bibitem[{{Green} {et~al.}(2009){Green}, {Caswell}, {Fuller}, {Avison},
  {Breen}, {Brooks}, {Burton}, {Chrysostomou}, {Cox}, {Diamond}, {Ellingsen},
  {Gray}, {Hoare}, {Masheder}, {McClure-Griffiths}, {Pestalozzi}, {Phillips},
  {Quinn}, {Thompson}, {Voronkov}, {Walsh}, {Ward-Thompson}, {Wong-McSweeney},
  {Yates}, \& {Cohen}}]{green+2009}
{Green}, J.~A., {Caswell}, J.~L., {Fuller}, G.~A., {et~al.} 2009, \mnras, 392,
  783

\bibitem[{{Green} {et~al.}(2010){Green}, {Caswell}, {Fuller}, {Avison},
  {Breen}, {Ellingsen}, {Gray}, {Pestalozzi}, {Quinn}, {Thompson}, \&
  {Voronkov}}]{green2010}
{Green}, J.~A., {Caswell}, J.~L., {Fuller}, G.~A., {et~al.} 2010, \mnras, 409,
  913

\bibitem[{{Green} {et~al.}(2012){Green}, {Caswell}, {Fuller}, {Avison},
  {Breen}, {Ellingsen}, {Gray}, {Pestalozzi}, {Quinn}, {Thompson}, \&
  {Voronkov}}]{green2012}
{Green}, J.~A., {Caswell}, J.~L., {Fuller}, G.~A., {et~al.} 2012, \mnras, 420,
  3108

\bibitem[{{G{\"u}sten} {et~al.}(2006){G{\"u}sten}, {Booth}, {Cesarsky},
  {Menten}, {Agurto}, {Anciaux}, {Azagra}, {Belitsky}, {Belloche}, {Bergman},
  {De Breuck}, {Comito}, {Dumke}, {Duran}, {Esch}, {Fluxa}, {Greve}, {Hafok},
  {H{\"a}upl}, {Helldner}, {Henseler}, {Heyminck}, {Johansson}, {Kasemann},
  {Klein}, {Korn}, {Kreysa}, {Kurz}, {Lapkin}, {Leurini}, {Lis}, {Lundgren},
  {Mac-Auliffe}, {Martinez}, {Melnick}, {Morris}, {Muders}, {Nyman}, {Olberg},
  {Olivares}, {Pantaleev}, {Patel}, {Pausch}, {Philipp}, {Philipps},
  {Sridharan}, {Polehampton}, {Reveret}, {Risacher}, {Roa}, {Sauer}, {Schilke},
  {Santana}, {Schneider}, {Sepulveda}, {Siringo}, {Spyromilio}, {Stenvers},
  {van der Tak}, {Torres}, {Vanzi}, {Vassilev}, {Weiss}, {Willmeroth},
  {Wunsch}, \& {Wyrowski}}]{guesten}
{G{\"u}sten}, R., {Booth}, R.~S., {Cesarsky}, C., {et~al.} 2006, in Presented
  at the Society of Photo-Optical Instrumentation Engineers (SPIE) Conference,
  Vol. 6267, Ground-based and Airborne Telescopes. Edited by Stepp, Larry M..
  Proceedings of the SPIE, Volume 6267, pp. 626714 (2006).

\bibitem[{{Hoare} \& {Franco}(2007)}]{hoare-2007}
{Hoare}, M.~G. \& {Franco}, J. 2007, ArXiv e-prints

\bibitem[{{Hoare} {et~al.}(2012){Hoare}, {Purcell}, {Churchwell}, {Diamond},
  {Cotton}, {Chandler}, {Smethurst}, {Kurtz}, {Mundy}, {Dougherty}, {Fender},
  {Fuller}, {Jackson}, {Garrington}, {Gledhill}, {Goldsmith}, {Lumsden},
  {Mart{\'{\i}}}, {Moore}, {Muxlow}, {Oudmaijer}, {Pandian}, {Paredes},
  {Shepherd}, {Spencer}, {Thompson}, {Umana}, {Urquhart}, \&
  {Zijlstra}}]{hoare2012}
{Hoare}, M.~G., {Purcell}, C.~R., {Churchwell}, E.~B., {et~al.} 2012, \pasp,
  124, 939

\bibitem[{{Holland} {et~al.}(1999){Holland}, {Robson}, {Gear}, {Cunningham},
  {Lightfoot}, {Jenness}, {Ivison}, {Stevens}, {Ade}, {Griffin}, {Duncan},
  {Murphy}, \& {Naylor}}]{Holland}
{Holland}, W.~S., {Robson}, E.~I., {Gear}, W.~K., {et~al.} 1999, \mnras, 303,
  659

\bibitem[{{Humphreys} \& {Larsen}(1995)}]{humphreys-1995}
{Humphreys}, R.~M. \& {Larsen}, J.~A. 1995, \aj, 110, 2183

\bibitem[{{Joint Iras Science}(1994)}]{iras_catalog}
{Joint Iras Science}, W.~G. 1994, VizieR Online Data Catalog, 2125, 0

\bibitem[{{Krumholz} \& {Bonnell}(2007)}]{krumholz-2007}
{Krumholz}, M.~R. \& {Bonnell}, I.~A. 2007, ArXiv e-prints

\bibitem[{{Lawrence} {et~al.}(2007){Lawrence}, {Warren}, {Almaini}, {Edge},
  {Hambly}, {Jameson}, {Lucas}, {Casali}, {Adamson}, {Dye}, {Emerson},
  {Foucaud}, {Hewett}, {Hirst}, {Hodgkin}, {Irwin}, {Lodieu}, {McMahon},
  {Simpson}, {Smail}, {Mortlock}, \& {Folger}}]{ukidss2007}
{Lawrence}, A., {Warren}, S.~J., {Almaini}, O., {et~al.} 2007, \mnras, 379,
  1599

\bibitem[{{McKee} \& {Ostriker}(2007)}]{mckee-ostriker-2007}
{McKee}, C.~F. \& {Ostriker}, E.~C. 2007, \araa, 45, 565

\bibitem[{{McKee} \& {Tan}(2002)}]{mckee-tan-2002}
{McKee}, C.~F. \& {Tan}, J.~C. 2002, \nat, 416, 59

\bibitem[{{McKee} \& {Tan}(2003)}]{mckee-tan-2003}
{McKee}, C.~F. \& {Tan}, J.~C. 2003, \apj, 585, 850

\bibitem[{{Menten}(1991)}]{menten1991}
{Menten}, K.~M. 1991, \apjl, 380, L75

\bibitem[{{Molinari} {et~al.}(2010){Molinari}, {Swinyard}, {Bally}, {Barlow},
  {Bernard}, {Martin}, {Moore}, {Noriega-Crespo}, {Plume}, {Testi}, {Zavagno},
  {Abergel}, {Ali}, {Andr{\'e}}, {Baluteau}, {Benedettini}, {Bern{\'e}},
  {Billot}, {Blommaert}, {Bontemps}, {Boulanger}, {Brand}, {Brunt}, {Burton},
  {Campeggio}, {Carey}, {Caselli}, {Cesaroni}, {Cernicharo}, {Chakrabarti},
  {Chrysostomou}, {Codella}, {Cohen}, {Compiegne}, {Davis}, {de Bernardis}, {de
  Gasperis}, {Di Francesco}, {di Giorgio}, {Elia}, {Faustini}, {Fischera},
  {Fukui}, {Fuller}, {Ganga}, {Garcia-Lario}, {Giard}, {Giardino}, {Glenn},
  {Goldsmith}, {Griffin}, {Hoare}, {Huang}, {Jiang}, {Joblin}, {Joncas},
  {Juvela}, {Kirk}, {Lagache}, {Li}, {Lim}, {Lord}, {Lucas}, {Maiolo},
  {Marengo}, {Marshall}, {Masi}, {Massi}, {Matsuura}, {Meny}, {Minier},
  {Miville-Desch{\^e}nes}, {Montier}, {Motte}, {M{\"u}ller}, {Natoli}, {Neves},
  {Olmi}, {Paladini}, {Paradis}, {Pestalozzi}, {Pezzuto}, {Piacentini},
  {Pomar{\`e}s}, {Popescu}, {Reach}, {Richer}, {Ristorcelli}, {Roy}, {Royer},
  {Russeil}, {Saraceno}, {Sauvage}, {Schilke}, {Schneider-Bontemps},
  {Schuller}, {Schultz}, {Shepherd}, {Sibthorpe}, {Smith}, {Smith},
  {Spinoglio}, {Stamatellos}, {Strafella}, {Stringfellow}, {Sturm}, {Taylor},
  {Thompson}, {Tuffs}, {Umana}, {Valenziano}, {Vavrek}, {Viti}, {Waelkens},
  {Ward-Thompson}, {White}, {Wyrowski}, {Yorke}, \& {Zhang}}]{higal}
{Molinari}, S., {Swinyard}, B., {Bally}, J., {et~al.} 2010, \pasp, 122, 314

\bibitem[{{Peretto} \& {Fuller}(2009)}]{ref-peretto}
{Peretto}, N. \& {Fuller}, G.~A. 2009, \aap, 505, 405

\bibitem[{{Planck Collaboration} {et~al.}(2011){Planck Collaboration}, {Ade},
  {Aghanim}, {Arnaud}, {Ashdown}, {Aumont}, {Baccigalupi}, {Balbi}, {Banday},
  {Barreiro}, \& et~al.}]{planck2011}
{Planck Collaboration}, {Ade}, P.~A.~R., {Aghanim}, N., {et~al.} 2011, \aap,
  536, A7

\bibitem[{{Price} {et~al.}(2001){Price}, {Egan}, {Carey}, {Mizuno}, \&
  {Kuchar}}]{price2001}
{Price}, S.~D., {Egan}, M.~P., {Carey}, S.~J., {Mizuno}, D.~R., \& {Kuchar},
  T.~A. 2001, \aj, 121, 2819

\bibitem[{{Rosolowsky} {et~al.}(2010){Rosolowsky}, {Dunham}, {Ginsburg},
  {Bradley}, {Aguirre}, {Bally}, {Battersby}, {Cyganowski}, {Dowell},
  {Drosback}, {Evans}, {Glenn}, {Harvey}, {Stringfellow}, {Walawender}, \&
  {Williams}}]{rosolowsky2010}
{Rosolowsky}, E., {Dunham}, M.~K., {Ginsburg}, A., {et~al.} 2010, \apjs, 188,
  123

\bibitem[{{Saito} {et~al.}(2012){Saito}, {Hempel}, {Minniti}, {Lucas},
  {Rejkuba}, {Toledo}, {Gonzalez}, {Alonso-Garc{\'{\i}}a}, {Irwin},
  {Gonzalez-Solares}, {Hodgkin}, {Lewis}, {Cross}, {Ivanov}, {Kerins},
  {Emerson}, {Soto}, {Am{\^o}res}, {Gurovich}, {D{\'e}k{\'a}ny}, {Angeloni},
  {Beamin}, {Catelan}, {Padilla}, {Zoccali}, {Pietrukowicz}, {Moni Bidin},
  {Mauro}, {Geisler}, {Folkes}, {Sale}, {Borissova}, {Kurtev}, {Ahumada},
  {Alonso}, {Adamson}, {Arias}, {Bandyopadhyay}, {Barb{\'a}}, {Barbuy},
  {Baume}, {Bedin}, {Bellini}, {Benjamin}, {Bica}, {Bonatto}, {Bronfman},
  {Carraro}, {Chen{\`e}}, {Clari{\'a}}, {Clarke}, {Contreras}, {Corvill{\'o}n},
  {de Grijs}, {Dias}, {Drew}, {Fari{\~n}a}, {Feinstein},
  {Fern{\'a}ndez-Laj{\'u}s}, {Gamen}, {Gieren}, {Goldman},
  {Gonz{\'a}lez-Fern{\'a}ndez}, {Grand}, {Gunthardt}, {Hambly}, {Hanson},
  {He{\l}miniak}, {Hoare}, {Huckvale}, {Jord{\'a}n}, {Kinemuchi}, {Longmore},
  {L{\'o}pez-Corredoira}, {Maccarone}, {Majaess}, {Mart{\'{\i}}n}, {Masetti},
  {Mennickent}, {Mirabel}, {Monaco}, {Morelli}, {Motta}, {Palma}, {Parisi},
  {Parker}, {Pe{\~n}aloza}, {Pietrzy{\'n}ski}, {Pignata}, {Popescu}, {Read},
  {Rojas}, {Roman-Lopes}, {Ruiz}, {Saviane}, {Schreiber}, {Schr{\"o}der},
  {Sharma}, {Smith}, {Sodr{\'e}}, {Stead}, {Stephens}, {Tamura}, {Tappert},
  {Thompson}, {Valenti}, {Vanzi}, {Walton}, {Weidmann}, \&
  {Zijlstra}}]{vvv2012}
{Saito}, R.~K., {Hempel}, M., {Minniti}, D., {et~al.} 2012, \aap, 537, A107

\bibitem[{{Schuller}(2012)}]{Schuller2012}
{Schuller}, F. 2012, Proceeding of the SPIE conference, submitted

\bibitem[{{Schuller} {et~al.}(2009){Schuller}, {Menten}, {Contreras},
  {Wyrowski}, {Schilke}, {Bronfman}, {Henning}, {Walmsley}, {Beuther},
  {Bontemps}, {Cesaroni}, {Deharveng}, {Garay}, {Herpin}, {Lefloch}, {Linz},
  {Mardones}, {Minier}, {Molinari}, {Motte}, {Nyman}, {Reveret}, {Risacher},
  {Russeil}, {Schneider}, {Testi}, {Troost}, {Vasyunina}, {Wienen}, {Zavagno},
  {Kovacs}, {Kreysa}, {Siringo}, \& {Wei{\ss}}}]{Schuller}
{Schuller}, F., {Menten}, K.~M., {Contreras}, Y., {et~al.} 2009, \aap, 504, 415

\bibitem[{{Siringo} {et~al.}(2009){Siringo}, {Kreysa}, {Kov{\'a}cs},
  {Schuller}, {Wei{\ss}}, {Esch}, {Gem{\"u}nd}, {Jethava}, {Lundershausen},
  {Colin}, {G{\"u}sten}, {Menten}, {Beelen}, {Bertoldi}, {Beeman}, \&
  {Haller}}]{siringo-2009}
{Siringo}, G., {Kreysa}, E., {Kov{\'a}cs}, A., {et~al.} 2009, \aap, 497, 945

\bibitem[{{Tackenberg} {et~al.}(2012){Tackenberg}, {Beuther}, {Henning},
  {Schuller}, {Wienen}, {Motte}, {Wyrowski}, {Bontemps}, {Bronfman}, {Menten},
  {Testi}, \& {Lefloch}}]{Tackenberg-2012}
{Tackenberg}, J., {Beuther}, H., {Henning}, T., {et~al.} 2012, \aap, 540, A113

\bibitem[{{Thompson} {et~al.}(2006){Thompson}, {Hatchell}, {Walsh},
  {MacDonald}, \& {Millar}}]{scuba_thompson}
{Thompson}, M.~A., {Hatchell}, J., {Walsh}, A.~J., {MacDonald}, G.~H., \&
  {Millar}, T.~J. 2006, \aap, 453, 1003

\bibitem[{{Urquhart} {et~al.}(2008){Urquhart}, {Hoare}, {Lumsden}, {Oudmaijer},
  \& {Moore}}]{urquhart2007c}
{Urquhart}, J.~S., {Hoare}, M.~G., {Lumsden}, S.~L., {Oudmaijer}, R.~D., \&
  {Moore}, T.~J.~T. 2008, in Astronomical Society of the Pacific Conference
  Series, Vol. 387, Massive Star Formation: Observations Confront Theory, ed.
  H.~{Beuther}, H.~{Linz}, \& T.~{Henning}, 381

\bibitem[{{Urquhart} {et~al.}(2012){Urquhart}, {Moore}, \&
  {Schuller}}]{Urquhart-2012}
{Urquhart}, J.~S., {Moore}, T.~J.~T., \& {Schuller}, F. 2012, \mnras, Submitted

\bibitem[{{Vasyunina} {et~al.}(2009){Vasyunina}, {Linz}, {Henning}, {Stecklum},
  {Klose}, \& {Nyman}}]{Vasyunina2009}
{Vasyunina}, T., {Linz}, H., {Henning}, T., {et~al.} 2009, \aap, 499, 149

\bibitem[{{Wienen} {et~al.}(2012){Wienen}, {Wyrowski}, {Schuller}, {Menten},
  {Walmsley}, {Bronfman}, \& {Motte}}]{wienen2012}
{Wienen}, M., {Wyrowski}, F., {Schuller}, F., {et~al.} 2012, \aap, 544, A146

\bibitem[{{Wilcock} {et~al.}(2012){Wilcock}, {Ward-Thompson}, {Kirk},
  {Stamatellos}, {Whitworth}, {Elia}, {Fuller}, {DiGiorgio}, {Griffin},
  {Molinari}, {Martin}, {Mottram}, {Peretto}, {Pestalozzi}, {Schisano},
  {Plume}, {Smith}, \& {Thompson}}]{wilcock2012}
{Wilcock}, L.~A., {Ward-Thompson}, D., {Kirk}, J.~M., {et~al.} 2012, \mnras,
  422, 1071

\bibitem[{{Wyrowski}(2008)}]{wyrowski-2007}
{Wyrowski}, F. 2008, in Astronomical Society of the Pacific Conference Series,
  Vol. 387, Massive Star Formation: Observations Confront Theory, ed.
  {H.~Beuther, H.~Linz, \& T.~Henning}, 3

\end{thebibliography}

\end{document}